\documentclass[aps,pra,eps2pdf,twocolumn,superscriptaddress]{revtex4-2}

\usepackage[utf8]{inputenc}
\usepackage{graphicx}
\graphicspath{{../}}

\pdfoutput=1 

\usepackage[ruled,vlined]{algorithm2e}

\usepackage{dsfont}
\usepackage{amsmath}
\usepackage{amssymb,bm}
\usepackage{amsthm}
\usepackage{mathtools}
\usepackage{url}
\usepackage{microtype}
\usepackage{array, makecell}
\usepackage{boldline}
\usepackage{enumitem}

\usepackage{color,dsfont}
\usepackage{ragged2e}
\usepackage[colorlinks=true, hyperindex, breaklinks, linkcolor=blue, urlcolor=blue, citecolor=blue]{hyperref} % Physical Review style
\usepackage[normalem]{ulem}
\usepackage[capitalise]{cleveref}
\usepackage{mathrsfs}
\usepackage[caption=false]{subfig}
\usepackage{float}
\usepackage{verbatim}
\usepackage{latexsym}
\usepackage{setspace}
\usepackage{amsfonts}
\usepackage{stmaryrd}
\usepackage{xcolor}
\usepackage{enumitem}
\usepackage{lipsum}
\usepackage{multirow}
\AtBeginDocument{\tabcolsep=6pt}
\usepackage{booktabs}

\usepackage{graphicx} 
\usepackage{amsfonts}
\usepackage{amssymb}
\usepackage{mathrsfs}
\usepackage{theorem}
\usepackage{amsmath}
\usepackage{times}
\usepackage{color}
\usepackage{soul}

\usepackage{ifpdf}
\ifpdf
\usepackage{epstopdf}   
\fi
\usepackage{cleveref}

\newcommand{\ket}[1]{\ensuremath{\vert#1\rangle}}

\newcommand{\kb}[2]{\ensuremath{\vert #1 \rangle \langle #2 \vert}}

 \newcommand{\CC}[1]{\textcolor{black} {#1}} % Chris

\def\id{\mbox{\small 1} \!\! \mbox{1}}

\def\id{{\mathchoice {\rm 1\mskip-4mu l} {\rm 1\mskip-4mu l} {\rm 1\mskip-4.5mu l} {\rm 1\mskip-5mu l}}}

\begin{document}

\title{Universal quantum computing with twist-free and temporally encoded lattice surgery}
\author{Christopher Chamberland}
\affiliation{AWS Center for Quantum Computing, Pasadena, CA 91125, USA}
\affiliation{IQIM, California Institute of Technology, Pasadena, CA 91125, USA}
\author{Earl T.\ Campbell}
\affiliation{AWS Center for Quantum Computing, Cambridge, UK}

\begin{abstract}

Lattice surgery protocols allow for the efficient implementation of universal gate sets with two-dimensional topological codes where qubits are constrained to interact with one another locally. In this work, we first introduce a decoder capable of correcting spacelike and timelike errors during lattice surgery protocols. Afterwards, we compute logical failure rates of a lattice surgery protocol for a biased circuit-level noise model. We then provide a new protocol for performing twist-free lattice surgery, where we avoid twist defects in the bulk of the lattice. Our twist-free protocol eliminates the extra circuit components and gate scheduling complexities associated with the measurement of higher weight stabilizers when using twist defects. We also provide a protocol for temporally encoded lattice surgery that can be used to reduce both runtimes and the total space-time costs of quantum algorithms. Lastly, we propose a layout for a quantum processor that is more efficient for rectangular surface codes exploiting noise bias, and which is compatible with the other techniques mentioned above. 

%This document include: techniques for decoding surface codes during lattice surgery; a protocol for twist-free lattice surgery; a protocol for temporally encoded lattice surgery that can improved both runtimes and total space-time algorithm costs; and a proposed quantum computing layout that is more efficient for biased noise surface codes and compatible with the other techniques described in this document.
\end{abstract}

\maketitle

%Architectures for fault-tolerant quantum computation protect qubits inside quantum error corrections codes and allowing for implementation of logic gates without leaving the protection of the code

\section{Introduction}
\label{sec:Intro}

Fault-tolerant quantum computing architectures enable the protection of logical qubits from errors by encoding them in error correcting codes, while simultaneously allowing for gates to be performed on such qubits. Importantly, failures arising during the implementation of logical gates do not result in uncorrectable errors as long as the total number of such failures remains below a certain fraction of the code distance~\cite{preskill1998reliable,terhal2015quantum,campbell2017roads,kitaev2003fault,Fowler12}.  In most practical settings, quantum logic gates are split into two categories. The first category corresponds to Clifford operations, which can be efficiently simulated by classical computers. The second category corresponds to non-Clifford operations, which cannot be efficiently simulated using purely classical resources.  Early proposals for fault-tolerant quantum computation used transversal gates to perform logical Clifford operations~\cite{shor1996fault}.  Later, it was shown that by braiding defects in a surface code, some Clifford operations could be realised fault-tolerantly in a 2D local architecture with a high-threshold~\cite{raussendorf2007topological}. Recently, lattice surgery~\cite{horsman2012surface} has replaced the braiding approach due to its ability to retain locality constraints and high thresholds (features which are required by many hardware architectures), while additionally offering a much lower resource cost~\cite{fowler2018low,BrownPokingHoles,litinski2018lattice,litinski2019game}.  These approaches all perform non-Clifford gates by teleportation~\cite{gottesman1999demonstrating,zhou2000methodology} of magic states prepared by some distillation procedure~\cite{BraKit05,Bravyi12,Knill12,Jones13,campbell2018magic,ChamberlandMagic,chamberland2020very}.  Alternative ideas have been proposed for circumventing the need for magic states~\cite{bombin2006topological,jochym2014using,bombin2016dimensional,Yoder2016}, but detailed studies~\cite{ChamberlandPRL,chamberland2017overhead,beverland2021cost} have not found any of these alternatives to be competitive for a wide range of failure rates below the surface code threshold.

% Throughout the history of the field there have been a sequence of ideas where Clifford operations are realised by some low-overhead means

% Early proposals for fault-tolerant logic used a combination of transversal gates for Clifford operators and teleportation for non-Clifford gates.

% include both an approach to error correction and an approach to performing logic without leaving the protection of an error correction code.  

Our work introduces the following key results. After briefly reviewing the model of Pauli based computation and its implementation via lattice surgery in \cref{sec:BriefReview}, we then explicitly provide a decoder compatible with lattice surgery in \cref{sec:LatticeSurgeryDecoder}. In particular, our decoder is capable of correcting both spacelike and timelike errors that occur during lattice surgery protocols. We then perform simulations of an $X \otimes X$ Pauli measurement using a biased circuit-level noise model.

In \cref{sec:TwistFreeLattice}, we introduce a twist-free approach for measuring arbitrary Pauli operators using the surface code. Our approach avoids the extra circuit and gate scheduling complexities that arise when using twists, where by twists we refer to lattices which contain twist defects in the bulk (see for instance \cref{fig:Twist} in \cref{subsec:RevLatticeSurgery}). We show that the approximate cost of avoiding twists is a $2 \times$ slowdown in the algorithm runtime and a negligibly small additive cost to the number of logical qubits. We expect that twist-based lattice surgery has it own associated costs, which may exceed those of our twist-free approach, but twist performance has never been fully quantified and so represents a currently unknown factor in quantum computing design.

In \cref{sec:FastLatticeSurgery}, we show how to reduce algorithm runtimes using a new technique that we call temporal encoding of lattice surgery. By using fast lattice surgery operations (which are inevitably noisier), errors arising from the extra noise can be corrected by encoding the sequence of measured Pauli operators within a classical error correcting code. The resulting runtime improvement grows (as a multiplicative factor) with the parallelizability of the algorithm and total algorithm runtime. We find that in a regime of interest to quantum algorithms of a practical scale, we can achieve a $2.2\times$ runtime improvement. Our temporal encoding does not directly lead to additional qubit overhead costs since it occurs in the time domain, and so the overall spacetime complexity is improved.

Lastly, in \cref {sec:routing}, we describe our core-cache architecture. We show that by using thin rectangular strips of surface codes for settings where a large noise bias is present, routing overhead costs in our proposed architecture add a factor of $1.5 \times$ to the total resource costs for performing lattice surgery. This can be compared with the $2 \times$ cost of Litinski's fast data access structures~\cite{litinski2019game}. Furthermore, we provide a layout that compactly stores surface code patches in a cache to further reduce the extra overhead arising from routing costs, at the cost of some additional time needed for reading/writing to the cache. Using the numerical results obtained in \cref{sec:LatticeSurgeryDecoder}, in \cref{App:OverheadCalcs} we provide resource cost estimates for simulating the Hubbard model using our core-cache architecture. 

\section{Brief review of universal quantum computing via lattice surgery}
\label{sec:BriefReview}

In \cref{subsec:RevPauliComp}, we briefly review the principles of Pauli based computation used throughout this work. We then review in \cref{subsec:RevLatticeSurgery} how multi-qubit Pauli operators are measured using lattice surgery.

\subsection{Overview of Pauli based computation}
\label{subsec:RevPauliComp}

\begin{figure}
    \centering
    \includegraphics[width=200pt]{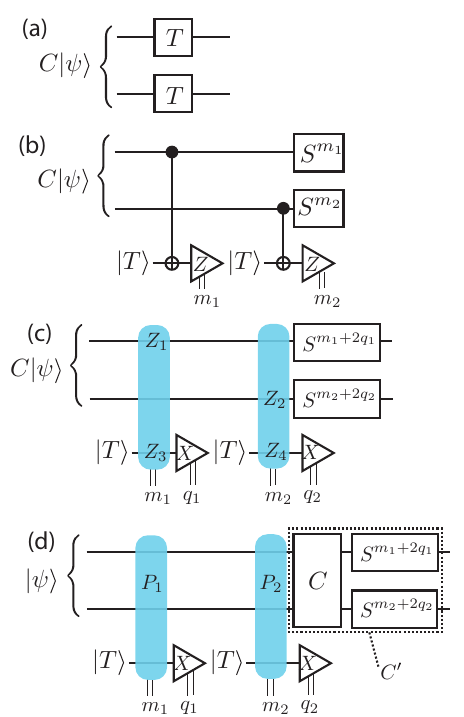}
    \caption{Equivalent approaches to implementing two $T$ gates. Blue rounded rectangles show multi-qubit Pauli measurements. (a) A simple unitary circuit approach.  (b) Using gate teleportation with $\vert T \rangle := (\vert 0 \rangle + e^{i \pi /4} \vert 1 \rangle)/\sqrt{2}$ magic states, CNOT gates, Pauli $Z$ measurements with outcomes $m_1$ and $m_2$ and classical conditioned Cliffords based on these outcomes.  The conditional $S$ gates are $S=\kb{0}{0} + i \kb{1}{1}$. (c) Using gate teleportation with the CNOT gates replaced by two-qubit Pauli measurements. (d) Given an input state $\vert \psi \rangle$ which carries a Clifford frame correction $C$, we have conjugated $C$ through the circuit so that the multi-qubit Pauli-measurements are now $P_1 =C Z_1 Z_3 C^\dagger$ and $P_2 =C Z_2 Z_4 C^\dagger$ (which commute) and the output state carries a new Clifford frame correction $C' =S_1^{m_1+ 2 q_1} S_2^{m_2+ 2 q_2} C $. This last circuit represents the standard PBC approach for computing $T^{\otimes 2}$.}
    \label{fig:TwoTgates}
\end{figure}

In the model of Pauli-based computation (PBC), we have a reserve of magic states and drive the computation by performing a sequence of multi-qubit Pauli measurements $\{ P_1, P_2, \ldots , P_\mu \}$ where later Pauli measurements depend on measurement outcomes of earlier measurements. In this notation, $P_2$ does not denote a specific Pauli, but one conditional on the outcome of $P_1$.  This conditionality occurs because (in the circuit picture) each Pauli measurement would be followed by a conditional Clifford operation. However, in a PBC, these Cliffords are conjugated to the end of the computation, thereby changing subsequent Pauli measurements.  Since in a PBC all Cliffords are performed ``in software", it is clear the algorithm runtime will be independent of the Clifford complexity.  The idea of PBC appears throughout the literature, but the phrase Pauli-based computation was first coined in Ref.~\cite{bravyi2016trading}. In \cref{fig:TwoTgates}, we present several computationally equivalent circuit diagrams for performing 2 $T$ gates, with the last diagram representing the PBC approach.

In \cref{subsec:RevLatticeSurgery}, we review how multi-qubit Pauli measurements can be performed using lattice surgery. Crucially, even when Pauli operators commute, it might not be possible to measure such Pauli operators simultaneously due to the extra space required to perform lattice surgery (known as the routing space). We can be obstructed from measuring commuting Pauli operators when the required lattice surgery operations need access to the same routing space.  Therefore, it is appropriate to consider sequentially measuring each Pauli operator, which we call a sequential Pauli based computation (seqPBC). In seqPBC, the time required to execute all the Pauli measurements will then be proportional to $T_{\mathrm{PBC}}=(d_m+1) \mu$ where we budget $+1$ for resetting qubits between lattice surgery operations. Here $d_m$ corresponds to the number of rounds of stabilizer measurements during lattice surgery, and $\mu$ is the number of sequential Pauli operators being measured. The proportionality factor will depend on the time required to measure the surface code stabilizers during one syndrome measurement round.  

\begin{figure*}
    \centering
    \includegraphics[width=400pt]{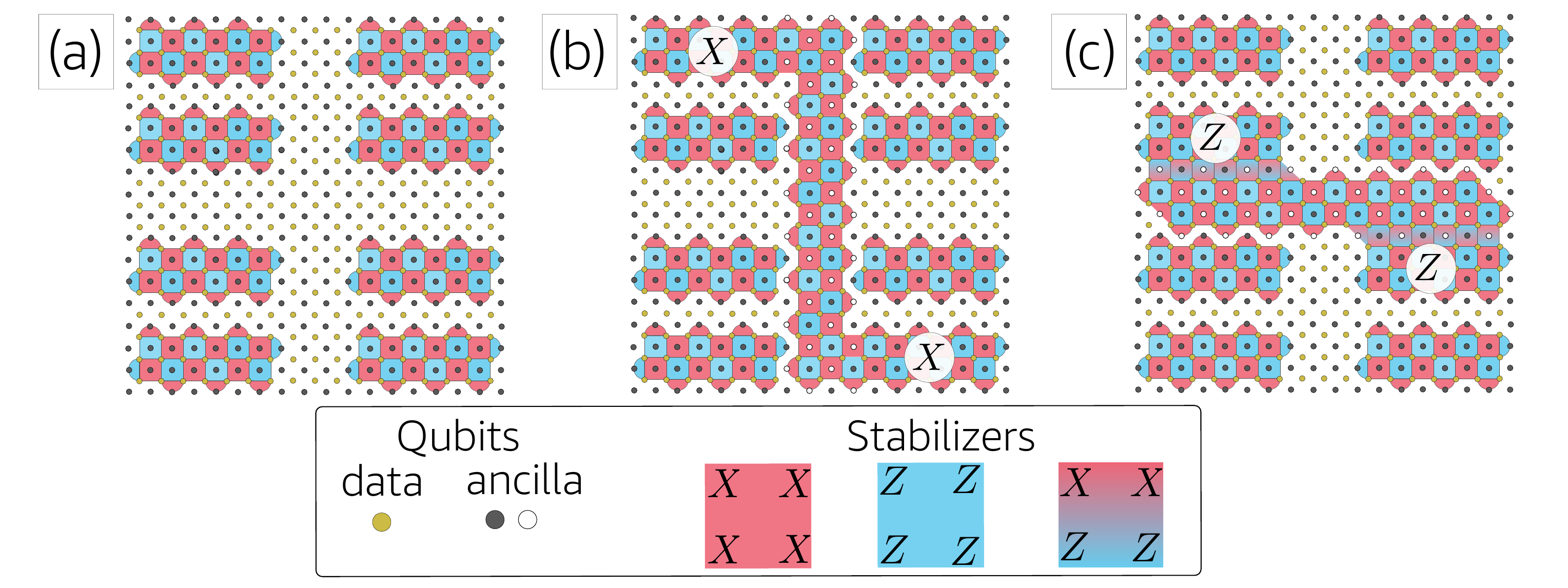}
    \caption{A simple example of lattice surgery illustrating the space overhead needed for routing.  In (a), we show 8 rectangular surface code patches with some surrounding idle qubits that form the routing space. In (b) we show the stabilizers needed to measure a logical $X \otimes X$ operator between the separated surface code patches using the routing space between them. Similarly, in (c) we show the stabilizers needed to measure a logical $Z \otimes Z$ operator between the separated surface code patches using the available routing space. Stabilizers of mixed $X$ and $Z$-type are referred to as domain walls and are used to ensure that the $d_z$ distance of the surface code does not decrease when measuring $Z$-type logical Pauli operators via lattice surgery. White ancillas mark the stabilizers that directly contribute towards computing the parity of the $X \otimes X$ and $Z \otimes Z$ measurement outcomes.}
    \label{fig:SimpleSurgery}
\end{figure*}

There are several contributions to the number of Pauli measurements $\mu$.  At a high level of the stack, we may think of a quantum algorithm as consisting of a series of unitaries with some Pauli measurements for readout, and let $N_A$ denote the number of such algorithmic readout measurements. However, as we saw in \cref{fig:TwoTgates}, non-Clifford unitaries are performed by measurements.   If an algorithm has $N_T$ $T$-gates, then we also need an additional $N_T$ Pauli measurements.  The Clifford plus $T$ gate set is universal. However, it is advantageous to use an overcomplete gate set such as Clifford plus $T$ and Toffoli.  While Toffoli can be synthesized using 4 $T$ gates~\cite{jones13b}, it is often more efficient to directly prepare Toffoli magic states~\cite{jones13b,eastin13,gidney2019efficient,chamberland2020building}. Furthermore, it only takes 3 Pauli measurements to teleport a Toffoli state rather than the 4 measurements needed to teleport $T$ states and then synthesize a Toffoli.  As such, if an algorithm can be executed with $N_{\text{TOF}}$ Toffoli gates and $N_T$ $T$-gates, then we need $N_T + 3 N_{\text{TOF}}$ measurements to perform these teleportations.  Further refinements are possible by using an even richer gate set and preparing more exotic states~\cite{campbell2017unified,campbell2017unifying,haah2019codes}, but we will not explicitly discuss those schemes here.  Lastly, we can replace some non-Clifford gates with Pauli measurements and feedforward (with no magic state needed). For instance, such a measurement appears in Gidney's circuits for adders~\cite{gidney2018halving} and more generally any uncomputation subroutine of an algorithm where the Toffoli gates are replaced with Pauli measurements and feedforward.  We use $N_{\text{unTOF}}$ to denote the number of Toffoli uncomputations performed in this manner. Hence, for a Clifford plus $T$ and Toffoli gate set, the total number of Pauli measurements will be $\mu = N_A + N_T + 3 N_{\text{TOF}} + N_{\text{unTOF}}$. Note that in many algorithms, Toffolis exclusively appear in compute/uncompute pairs, and then we have $N_{\text{unTOF}}=N_{\text{TOF}}$.  Furthermore, algorithms often only have a small number of qubit readouts, so $N_A \ll N_T, N_{\text{TOF}}$. As such, we commonly have $\mu \approx N_T + 4 N_{\text{TOF}}$.

An architecture also requires time $T_{\mathrm{magic}}$ to produce the required magic states, say $N_T$ $T$-states and $N_{\text{TOF}}$ Toffoli states.  If an architecture produces all the required magic states in a shorter amount of time than is required to teleport them, so that $T_{\mathrm{magic}} < T_{\mathrm{PBC}}$, then we say the seqPBC is Clifford-bottle necked. On the other hand, if  $T_{\mathrm{magic}} \geq T_{\mathrm{PBC}}$, we say it is magic-state bottle necked.  The running time of the algorithm will be determined by $\mathrm{max}\{ T_{\mathrm{magic}} , T_{\mathrm{PBC}} \}$. 

It is informative to briefly review the impact of these bottle necks on the history of algorithm resource analysis. The time $T_{\mathrm{magic}}$ can be made arbitrarily small, by simply increasing the number of magic state factories, though this comes at an increased qubit cost. Some early algorithm overhead estimates~\cite{Fowler12,Ogorman16,campbell2019applying} minimized $T_{\mathrm{magic}}$ by having a large number of factories leading to widespread claims that magic state factories could be $\sim 99\%$ of the whole device.  However, such resource estimates ignore the question of how quickly these states can be teleported and ignored the $T_{\mathrm{PBC}}$ bottleneck.  Accounting for this bottleneck, there is no benefit from pushing $T_{\mathrm{magic}}$ to be small $T_{\mathrm{magic}} \ll T_{\mathrm{PBC}}$.  As such, more recent and careful overhead analyses~\cite{berry2019qubitization,berry2019qubitization,kivlichan2020improved,chamberland2020building,lee2021even} have assumed only a few factories, which is enough to achieve $T_{\mathrm{magic}} \approx T_{\mathrm{PBC}}$ and results in only a small percentage of device footprint (often less than $1\%$) being used as a magic state factory.  While these analyses have minimal qubit cost, the runtime now is $T_{\mathrm{PBC}}$ bottlenecked, motivating rigorous approaches to beating the  $T_{\mathrm{PBC}}$ bottleneck.  Later, in \cref{subsec:ParallelPaulis}, we will review prior lattice surgery methods to speed up algorithms by using additional teleportation gadgets, though this comes at a high qubit cost.  We will then introduce our own approach that instead uses temporal encoding of lattice surgery (TELS).

 \begin{figure*}
    \centering
    \includegraphics[width=420pt]{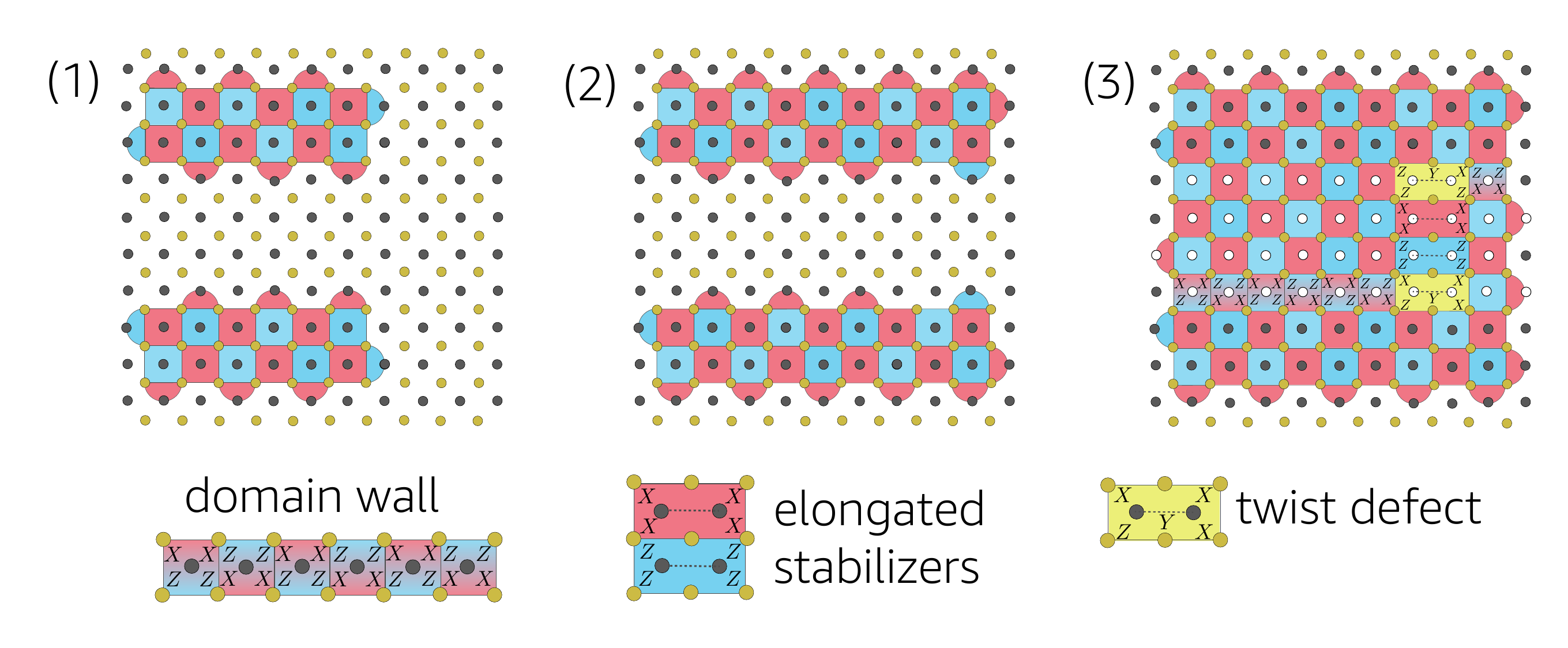}
    \caption{A simple example of a lattice surgery protocol using a twist defect in the bulk of the lattice to measure a $Y \otimes Y$ logical operator.  Step (1) is the initial setup of 2 surface code patches. In step (2), the surface code patches are extended and corners moved using the routing space. Such an extension allows logical $Y$ operators of the surface code to be expressed along a horizontal boundary. In step (3), we measure $Y \otimes Y$ using a combination of domain walls, elongated stabilizers and two twist defects. Domain walls measure $Z \otimes Z \otimes X \otimes X$ stabilizers and typically offer no additional challenge compared to normal stabilizer measurements.  Elongated stabilizers are long range operators that pose an additional difficulty in hardware implementations.   Elongated stabilizers can be implemented using either: two ancilla qubits (e.g. prepared in a GHZ state) that we connect with a dashed line; or these two ancilla qubits must be merged into a single qubit (hardwired in the architecture) and using  long-range gates. Twist defects (yellow plaquettes) present the biggest difficulty since in addition to the challenges faced by elongated stabilizers, they also require a weight-five measurement. }
    \label{fig:Twist}
\end{figure*}

\subsection{Overview of lattice surgery}
\label{subsec:RevLatticeSurgery}

For quantum hardware where physical qubits can only interact with one another locally, lattice surgery is a fault-tolerant protocol that can be used to measure arbitrary multi-qubit Pauli operators. The main idea is to encode the logical qubits in some topological code arranged in a two-dimensional layout (in this work, all logical qubits are encoded in the rotated surface code \cite{TomitaSvore}). The layout contains extra routing space between the surface code patches which consists of additional qubits. By applying the appropriate gauge-fixing operations in the routing space (see for instance Ref.\cite{Vuillot_2019}), which involves measuring surface code stabilizers, the surface code patches involved in the Pauli measurement are merged into one larger surface code patch. After gauge fixing, the parity of the measurement outcome of the multi-qubit Pauli operator being measured is obtained by taking the products of the appropriate stabilizers in the routing space. Lastly, the surface code patches for each logical qubit can be detached from the merged patch by measuring the qubits in the routing space in the appropriate basis. An illustration of $X \otimes X$ and $Z \otimes Z$ Pauli measurements is shown in \cref{fig:SimpleSurgery}. Products of the surface code stabilizers marked by white ancilla qubits give the parity of the $X \otimes X$ and $Z \otimes Z$ measurement outcomes. Note that for $Z$-type Pauli measurements, we use domain walls at the $Z$ logical boundaries of the surface code patches. Domain walls correspond to stabilizers of mixed $X$ and $Z$-type as illustrated in \cref{fig:Twist}. The primary reason for using domain walls is to prevent a reduction in minimum-weight representatives of logical $Z$ operators during lattice surgery.

To measure multi-qubit Pauli operators containing $Y$ terms, one option is to extend the surface code patches using the routing space in such a way that the logical $Y$ operators can be expressed along horizontal boundaries of the surface code. Logical $Y$ operators can then be measured using a twist defect as shown in \cref{fig:Twist}. Note that such a protocol requires measuring a weight-five operator, such as the ones shown in yellow in \cref{fig:Twist}. Such high-weight measurements can be undesirable for many hardware architectures. An alternative approach which does not require the extension of surface code patches and the use of twist-defects in the bulk is provided in \cref{sec:TwistFreeLattice}. 

When performing a lattice surgery measurement of a logical Pauli operator, there will be some probability that we obtain the wrong outcome.  Even with large code distances, the lattice surgery measurement could still fail due to time-like errors occurring during the finite time allowed for lattice surgery. The probability of these failure events is exponentially suppressed in the number of rounds $d_m$ for which we repeat the stabilizer measurements during lattice surgery. Therefore, we call $d_m$ the measurement distance which quantifies the protection against repeated measurement failures during lattice surgery, and hence explains our choice of subscript. Defining $d_z$ and $d_x$ to be minimum-weights of logical $Z$ and $X$-type operators of the surface code, this exponential suppression will hold until $d_m \gg O(d_z, d_x)$ when logical Pauli errors become the dominate mechanism again.  Let us assume that code distances $d_x$ and $d_z$ are chosen so that even a single logical $Z$ and $X$ error is very unlikely over the course of the whole computation.  We expect, and numerically find, that these timelike errors occur with a probability $\mathbb{P}$ for which we have a bound of the form
\begin{equation}  
 \mathbb{P} \leq  L a ( p b)^{c (d_m+1)} ,
 \label{eq:Perror}
\end{equation}
where $p$ quantifies the physical gate failure probabilities, $\{a,b,c\}$ are constants and $L$ is the area of the patch used for lattice surgery.  The value of $L$ will vary for different measurements and different layouts, but it will be convenient to think of it as a constant representing the worst (or average) case area of lattice surgery patches.

In general, if we want to sequentially perform $\mu$ Pauli measurements in the algorithm and we want them to fail with probability no more than $\delta$, then we choose $d_m$ to be large enough such that
\begin{equation}  \label{eq:FailureProbs}
   \mu L a ( p b)^{c (d_m+1)} \approx 1-(1- \mathbb{P} )^{\mu} \leq \delta
    \end{equation}
    where the approximation holds for small $\mathbb{P}$. In \cref{sec:LatticeSurgeryDecoder}, after introducing a decoder compatible with lattice surgery protocols, we will compute timelike failure probabilities in addition to probabilities for other noise processes given a biased circuit-level noise model. Such results will allow us to obtain accurate resource overhead estimates for implementing quantum algorithms and are discussed further in \cref{sec:routing}.

\section{Decoding timelike errors during lattice surgery}
\label{sec:LatticeSurgeryDecoder}

\begin{figure}%[t]
	\includegraphics[width=\columnwidth]{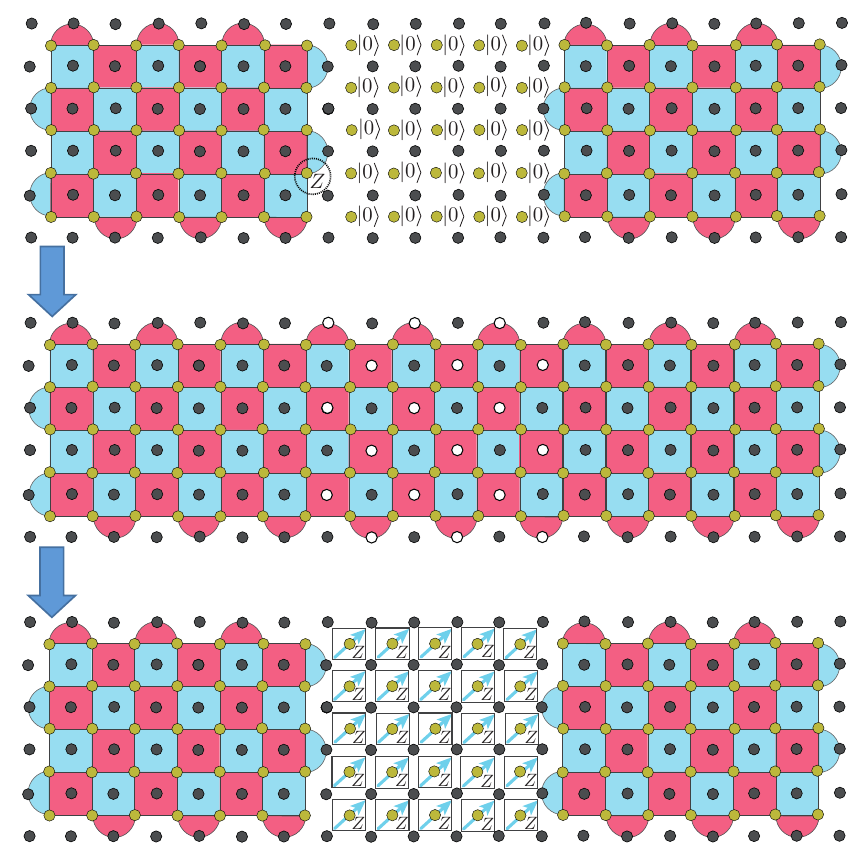}
	\caption{Example of an $X \otimes X$ measurement performed on two $d_x=5$, $d_z=7$ logical patches. Data qubits in the routing space are first prepared in the $\ket{0}$ state. In the first round of stabilizer measurements where the logical patches are merged into one large surface code patch, the product of all stabilizers marked by white vertices (which we call parity vertices) gives the result of the $X \otimes X$ measurement. After measuring the stabilizers of the merged patch for $d_m$ rounds, the patches are split by measuring the data qubits in the routing region in the $Z$ basis. In the first round of the merge, measurement errors occurring on parity vertices can result in a wrong $X \otimes X$ measurement outcome. Additionally, an odd number of data-qubit $Z$ errors along the boundary of the logical patches prior to the merge, such as the one circled in the top row of the figure, can also result in a wrong $X \otimes X$ measurement outcome.}
		\label{Fig:LatticeMerge}
\end{figure}

In this section, we provide an explicit decoding protocol for correcting both spacelike and timelike errors which can occur during lattice surgery protocols. In particular, our protocol protects logical qubits encoded in surface code patches while at the same time correcting logical multi-qubit Pauli measurement failures that can occur during lattice surgery. We then provide numerical results for performing $X \otimes X$ measurements showing both the logical multi-qubit Pauli measurement failure rate as a function of the number of syndrome measurement rounds and the logical qubit failure rates. 

\CC{The generalization of toric code decoders with periodic boundary conditions to the surface code was first done in Refs.~\cite{raussendorf2007topological,Rauss07,PhysRevA.80.052312} by adding virtual vertices at the boundaries of the surface code. Follow up work in Refs.~\cite{horsman2012surface,fowler2018low} provided a high-level account of how surface code decoders could be used in the context of lattice surgery. However, details such as the correct specification of boundary vertex locations for both spacelike and timelike failures were not provided. In \cref{subsec:DecodeAlgo}, our lattice surgery decoding algorithm makes use of boundary vertices for both spatial and timelike boundaries. Further, no} previous work has performed such realistic, circuit-level simulations of lattice surgery.  For comparison, in Ref.~\cite{chamberland2020building} timelike errors were simulated, but Ref.~\cite{chamberland2020building} used a toy model with idealized boundaries to exclude logical spacelike errors and thereby simplifying both the simulations and required decoding algorithm. In what follows, we will refer to a surface code patch encoding a logical qubit as \textit{a logical patch.} Space used for performing multi-qubit Pauli measurements via lattice surgery will be referred to as the \textit{routing space} or \textit{routing region}.

In \cref{Fig:LatticeMerge}, we provide an example of an $X \otimes X$ Pauli measurement performed between two $d_x=5$, $d_z=7$ logical patches. After preparing ancilla qubits (grey vertices) in the routing space in $\ket{+}$ and data qubits (yellow vertices) in $\ket{0}$, $X$-type surface code stabilizers are measured. The products of stabilizers marked by white vertices gives the measurement outcome of the logical $X \otimes X$ operator. In what follows, white vertices whose product gives the result of a $P_1 \otimes P_2 \otimes \cdots \otimes P_k$ Pauli measurement will be referred to as \textit{parity vertices}. We say that a logical timelike failure occurs if a set of errors result in the wrong parity measurement of $P_1 \otimes P_2 \otimes \cdots \otimes P_k$. Further, we assume that the logical patches are measured for $r$ rounds prior to being merged in round $r+1$. 

When measuring $P_1 \otimes P_2 \otimes \cdots \otimes P_k$ using lattice surgery, in addition to an odd number of measurement errors occurring in round $r+1$, an odd number of data-qubit errors along the boundaries of the logical patches prior to the merge can also result in a logical timelike failure. An example is provided in \cref{Fig:LatticeMerge} where a single data-qubit $Z$ error along a boundary of the left logical patch prior to the merge gives the wrong parity of $X \otimes X$. We also note that during the syndrome measurement round $r+1$, an odd number data-qubit errors which anticommute with $P_1 \otimes P_2 \otimes \cdots \otimes P_k$ will (unless corrected) also result in a logical timelike failure. 

The above examples show that in order to obtain the correct parity measurement of a $P_1 \otimes P_2 \otimes \cdots \otimes P_k$ operator in the presence of full circuit level noise, one must have a decoding scheme which, while constantly correcting errors on the logical patches, also corrects spacelike and timelike errors which can flip the parity of the measurement outcome. 

\begin{figure}%[th]
	\centering
	\subfloat[\label{fig:GraphsFuture}]{%
		\includegraphics[width=0.48\textwidth]{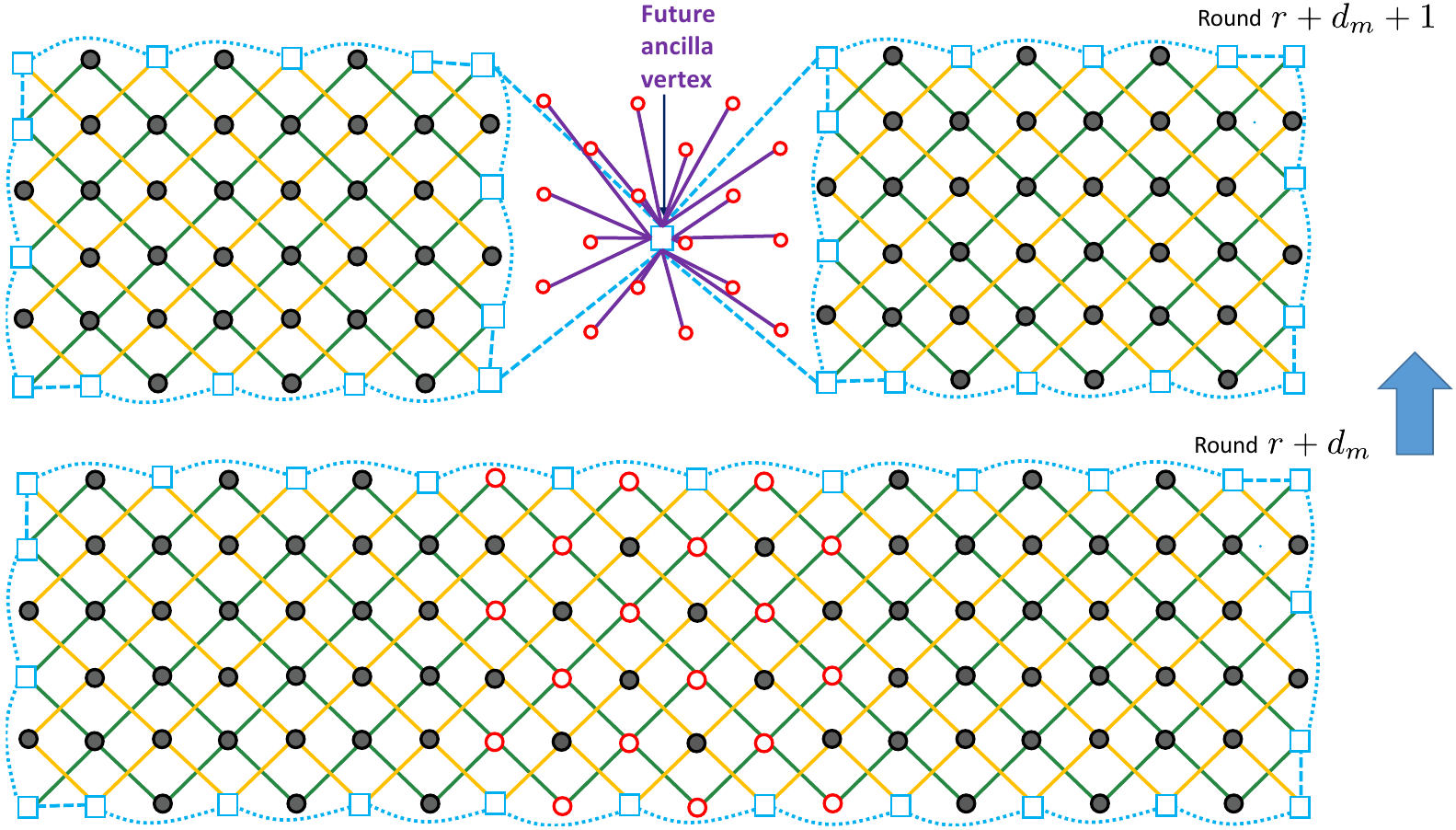}
	}
	\vfill
	\subfloat[\label{fig:GraphsPast}]{%
		\includegraphics[width=0.48\textwidth]{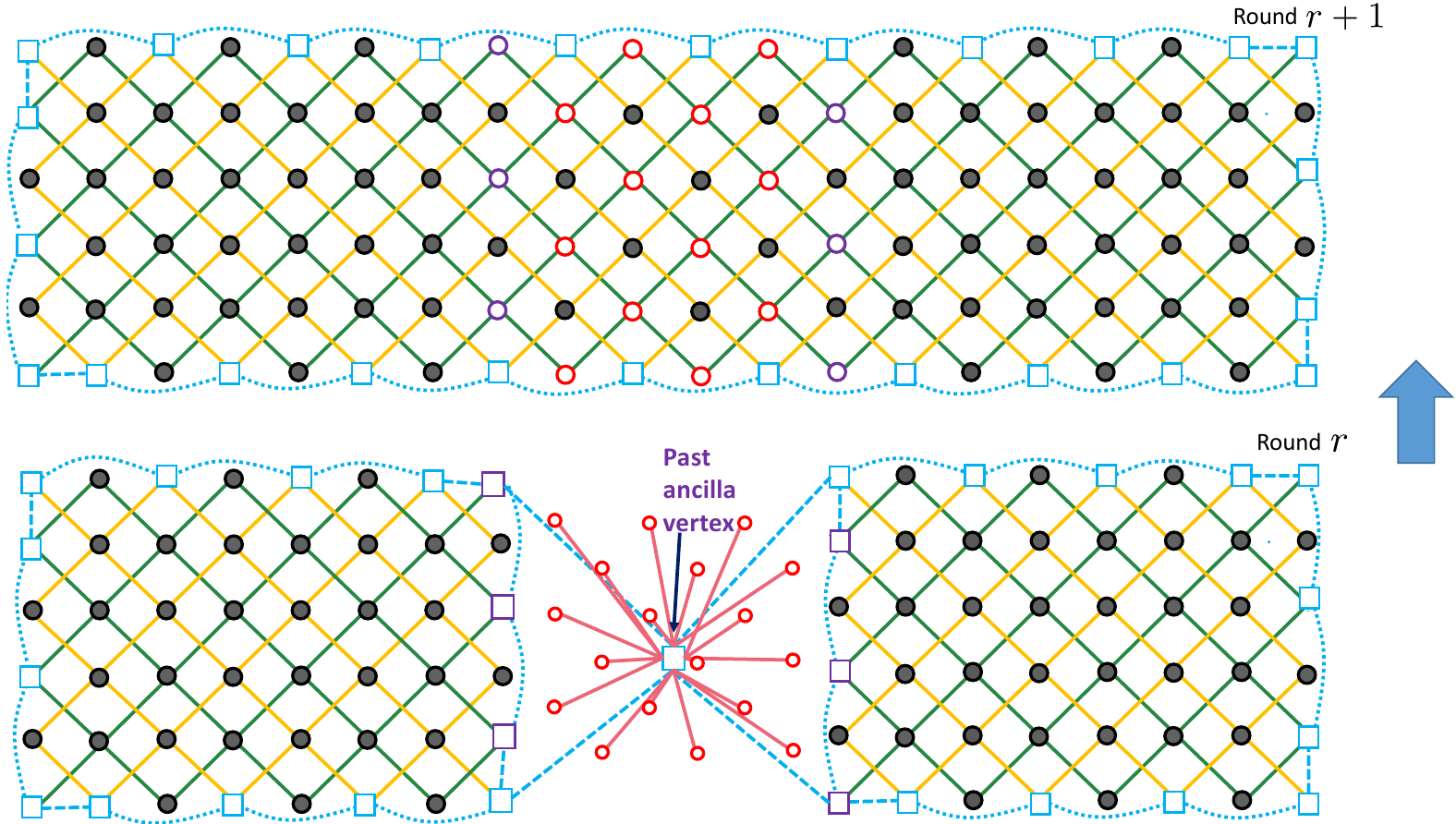}
	}
	\caption{Various two-dimensional slices of the matching graph used for performing an $X \otimes X$ Pauli measurement via lattice surgery. (a) Two-dimensional slices of the surface code matching graphs for syndrome measurement rounds $r+d_m$ (bottom) and $r+d_m+1$ (top). The graph in round $r+d_m+1$ includes the future ancilla vertex with future vertical edges (solid purple edges) connecting to the parity vertices (white vertices circled in red) of the matching graph in round $r+d_m$. (b) Two-dimensional slices of the surface code matching graphs for syndrome measurement rounds $r$ (bottom) and $r+1$ (top). The graph in round $r$ includes the past ancilla vertex with past vertical edges (solid pink edges) connecting to the parity vertices (white vertices circled in red) of the matching graph in round $r+1$. Transition vertices are the purple boundary vertices of the graph prior to the merge (note that transition vertices appear in rounds $1$ to $r$), in addition to the purple parity vertices in round $r+1$.}
	\label{fig:FullSurfaceGraphsWithTimeLike}
\end{figure}

\subsection{The decoding algorithm}
\label{subsec:DecodeAlgo}

In order to correct logical timelike failures using a minimum-weight-perfect-matching (MWPM) decoder~\cite{Edmonds65}, we must add timelike boundaries to the matching graphs of the surface code as shown in \cref{fig:FullSurfaceGraphsWithTimeLike}. In particular, we divide the measurement of an operator $P_1 \otimes P_2 \otimes \cdots \otimes P_k$ into three steps. In the first step, the logical patches are measured for $r$ rounds. In round $r + 1$, the patches are merged by measuring the appropriate operators in the routing space (see for instance \cref{Fig:LatticeMerge}) and the parity of the measurement outcome is given by the product of all parity vertices. The merged patches are measured for $d_m$ rounds, and then in round $d_m+1$, the qubits in the routing space are measured in the appropriate basis to split the patches back to their original configuration. In round $r$, we add extra virtual vertices to the matching graph with vertical edges which are incident to such vertices and to the parity vertices in round $r+1$ (see the pink edges in \cref{fig:GraphsPast} for the $X \otimes X$ measurement). We call such vertices \textit{past ancilla vertices} and the pink edges incident to them \textit{past vertical edges}. Similarly, in round $r+d_m+1$ (i.e. right after the split), we add virtual vertices to the matching graph with vertical edges which are incident to such vertices and to the parity vertices in round $r+d_m$ (see the purple edges in \cref{fig:GraphsFuture} for the $X \otimes X$ measurement). We call such vertices \textit{future ancilla vertices} and the purple edges incident to them \textit{future vertical edges}. Importantly, the pink vertical edges that are incident to the past ancilla vertices and to the parity vertices in round $r+1$ have zero weight, while the purple vertical edges incident to the parity vertices in round $r+d_m$ and the future ancilla vertices have non-zero weights. These weights are computed from all timelike failure processes which can result in measurement errors occurring in round $r + d_m$. When performing MWPM over the full syndrome history, the parity of the measurement outcome of $P_1 \otimes P_2 \otimes \cdots \otimes P_k$ is flipped if there are an odd number of highlighted vertical edges incident to parity vertices in rounds $r+1$ and $r+2$. In such a setting, one would require a sequence of consecutive measurement errors that is greater than $(d_m-1)/2$ in order to cause a timelike logical failure. Note that since the measurement outcomes of the parity vertices in round $r+1$ are random, such vertices are never highlighted. If a change in the measurement outcomes of a subset of the parity vertices are observed between rounds $r+1$ and $r+2$, then vertices corresponding to such parity vertices in round $r+2$ would be highlighted. Furthermore, green horizontal edges incident to the parity vertices in round $r+1$ are taken to have zero weight (or can be omitted). 

\begin{figure}%[t]
	\includegraphics[width=\columnwidth]{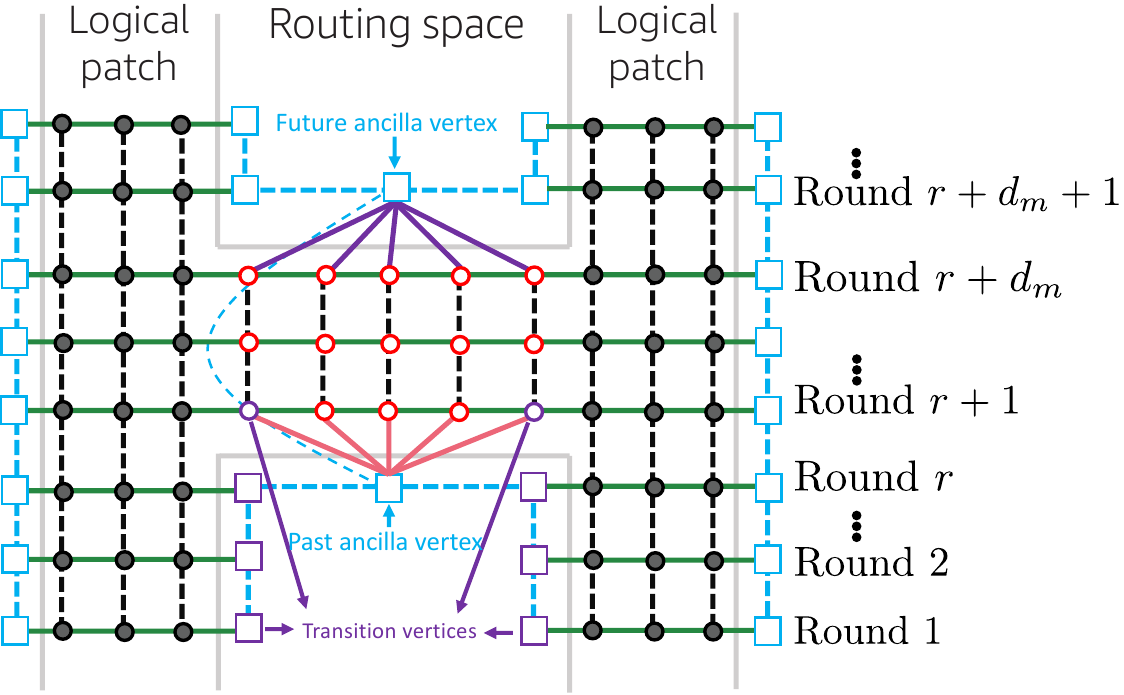}
	\caption{A two-dimensional slice of the surface code matching graph in the timelike direction for syndrome measurement rounds performed during an $X \otimes X$ measurement protocol using lattice surgery. The vertical axis corresponds to the timelike direction. We show a subset of the vertices and edges of the matching graph for correcting $Z$ errors for an $X \otimes X$ measurement using lattice surgery. Parity vertices are the white and red vertices in the ancilla patch region. Transition vertices are both boundary and parity vertices colored in purple. Pink edges are past vertical edges incident to the past ancilla vertex and parity vertices in round $r+1$, whereas purple edges are future ancilla edges incident to the parity vertices in round $r+d_m$ and the future ancilla vertex. We also add a dashed-blue weightless edge connecting the past and future ancilla vertices.}
		\label{Fig:Graphs2DSlice}
\end{figure}

\begin{algorithm}[t]
\SetAlgoLined
\KwResult{Data qubit and parity measurement corrections.}
 \textbf{initialize:} $v_{1} = v_{2} = 0$. Let $G_{r}$ be the graph for the surface code patches before, during and after the merge.\;
 \textbf{Measurement:} Measure the stabilizers of the split logical patches for $r$ rounds. Merge the logical patches in round $r+1$ via lattice surgery to perform the $P_1 \otimes P_2 \otimes \cdots \otimes P_k$ measurement and let $s_{\text{par}}$ be the parity of the measurement outcome. Repeat the stabilizer measurements of the merged patch for $d_m-1$ rounds.\; 
 \textbf{1)} Add the past ancilla vertex $v_{\text{past}}$ to $G_{r}$ for the round $r$ (round before the merge). Let $V^{(r+1)}_{\text{par}} = \{ v^{(1)}_{\text{par}}, \cdots, v^{(k)}_{\text{par}} \}$ be the set of parity vertices for the syndrome measurement round $r+1$. Add weightless past vertical edges to $G_{r}$ which are incident to $v_{\text{past}}$ and all vertices $v \in V^{(r+1)}_{\text{par}}$. Add weightless edges to $G_{r}$ which are between $v_{\text{past}}$ and virtual boundary edges of all surface code patches\;
 \textbf{2)} Add the future ancilla vertex $v_{\text{future}}$ to $G_{r}$ for the round $r + d_m + 1$ (round after the merge). Let $V^{(r+d_m)}_{\text{par}} = \{ \tilde{v}^{(1)}_{\text{par}}, \cdots, \tilde{v}^{(k)}_{\text{par}} \}$ be the set of parity vertices for the syndrome measurement round $r+d_m$. Add future vertical edges (of non-zero weight) to $G_{r}$ which are incident to $v_{\text{future}}$ and all vertices $v \in V^{(r+d_m)}_{\text{par}}$\;
 \textbf{3)} Add a weightless edge to $G_{r}$ which is incident to $v_{\text{past}}$ and $v_{\text{future}}$\;
 \textbf{4)} Set all edges incident to any two vertices $v_i, v_j \in V^{(r+1)}_{\text{par}}$ to have zero weight\;
 \textbf{5)} Given the full syndrome measurement history, if the total number of highlighted vertices (obtained by taking the difference between any two consecutive syndrome measurement rounds modulo 2) is odd, highlight $v_{\text{future}}$\;
 \textbf{6)} Implement MWPM on $G_{r}$. Set $v_1$ to be the number of highlighted edges incident to vertices in $V^{(r+1)}_{\text{par}}$ and $V^{(r+2)}_{\text{par}}$, and $v_2$ to be the number of highlighted edges incident to transition vertices in the data-qubit patch regions. If $v_1 + v_2$ is odd, set $s_{\text{par}} \rightarrow s_{\text{par}} + 1 \pmod{2}$\;
 \textbf{7)} Apply data qubit corrections based on all highlighted two-dimensional and space-time correlated edges.
 \caption{Decoding algorithm for measuring $P_1 \otimes P_2 \otimes \cdots \otimes P_k$ via lattice surgery.}
 \label{Alg:LatticeSurgery}
\end{algorithm}

\begin{figure*}%[th]
	\centering
	\subfloat[\label{fig:MarginalFail}]{%
		\includegraphics[width=0.48\textwidth]{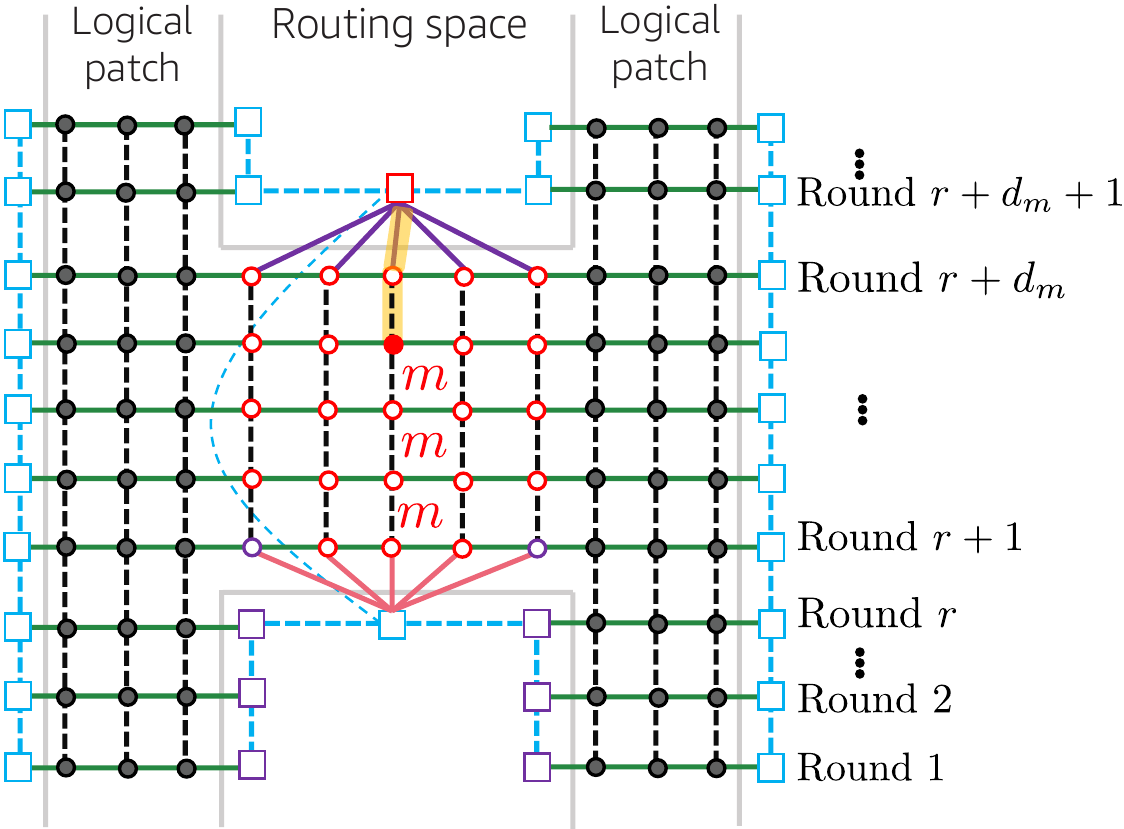}
	}
	%\vfill
	\subfloat[\label{fig:Fail110}]{%
		\includegraphics[width=0.48\textwidth]{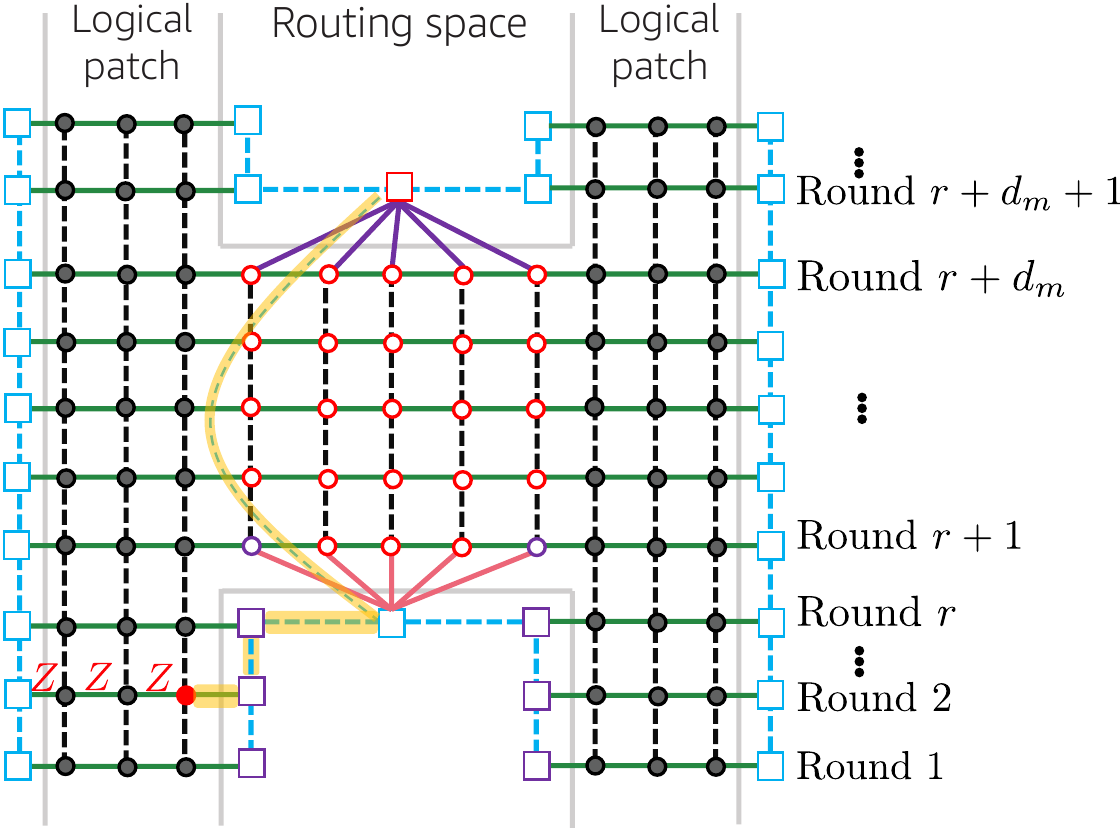}
	}
	\caption{(a) Example of a series of consecutive measurement errors on the same parity vertex located in the routing space resulting in a $(0,1,0,0)$ logical failure for an $X \otimes X$ measurement. (b) Example of a horizontal string of spacelike $Z$ data-qubit errors on the left logical patch resulting in a $(1,1,0,0)$ logical failure.}
	\label{fig:FailureMechanisms}
\end{figure*}

As mentioned above, a lattice surgery decoder also needs to correct logical timelike failures arising from sets of data qubit errors along boundaries of the logical qubit patches prior to merging them (recall the example shown in \cref{Fig:LatticeMerge}). The decoder also needs to correct wrong parity measurements arising from data qubit errors in round $r+1$ which anticommute with the Pauli operator being measured by lattice surgery. In constructing such a decoder, note that prior to merging the logical patches for the $X \otimes X$ measurement, a single $Z$ error along the relevant boundaries would result in a highlighted edge (after implementing MWPM over the full syndrome history) incident to one of the purple boundary vertices shown in \cref{fig:GraphsPast}. Note that such boundary vertices become parity vertices after merging the surface code patches. For the measurement of a general $P_1 \otimes P_2 \otimes \cdots \otimes P_k$ Pauli operator, we define \textit{transition vertices} to be vertices in the set $V^{(s)}_{Bd} = \{v^{(s)}_{b_1}, \cdots, v^{(s)}_{b_m} \}$, where $1 \le s \le r+1$. When $s < r+1$,  $\{v^{(s)}_{b_1}, \cdots, v^{(s)}_{b_m} \}$ are labels for boundary vertices of the graphs of split logical patches which become parity vertices in round $r+1$. If $s=r+1$, then $V^{(r+1)}_{Bd}$ is the set of parity vertices along the boundaries of the logical qubit patches and routing space used to merge the logical qubits (see for instance the parity vertices highlighted in purple in \cref{Fig:Graphs2DSlice}). Based on previous observations, after implementing MWPM over the full syndrome history of a multi-qubit Pauli measurement via lattice surgery, if an odd number of space-like highlighted edges are present in logical patches, and such edges are incident to transition vertices, the parity of the Pauli measurement needs to be flipped. An illustration of a two-dimensional slice of the matching graphs of \cref{fig:FullSurfaceGraphsWithTimeLike} in the timelike direction (which contains a subset of the spacelike edges and vertices) is shown in \cref{Fig:Graphs2DSlice}. In particular, the figure illustrates transition vertices, the past and future ancilla vertices in addition to the past and future vertical edges.

Combining all the notions introduced in this section, the decoding algorithm for implementing a multi-qubit Pauli measurement via lattice surgery is described in \cref{Alg:LatticeSurgery}. Each highlighted edge in step 7) of \cref{Alg:LatticeSurgery} encodes a particular data qubit correction. Writing such corrections as a binary row vector, where each column corresponds to a data qubit, we add all corrections arising from each highlighted edge using modulo-two arithmetic. Further, note that space-time correlated edges incident to parity vertices in round $r+1$ need to be treated with care in order to correct errors up to the full code distance. In particular, a subset of the space-time correlated edges incident to transition vertices can also contribute to $v_2$ (defined in \cref{Alg:LatticeSurgery}). A more careful treatment of such edges is provided in \cref{App:SpaceTime}.

\subsection{Decoder simplifications}

We point out a simplification that can be made in the implementation of \cref{Alg:LatticeSurgery}. Notice that vertices in $V^{(r+1)}_{\text{par}}$ are never highlighted during MWPM due to the random outcomes of stabilizers in the routing space marked by white vertices in round $r+1$. As such, one could remove all vertices in $V^{(r+1)}_{\text{par}}$ and instead have the past vertical edges incident to the vertices in $V^{(r+2)}_{\text{par}}$. In such a setting, edges incident to $V^{(r+1)}_{\text{par}}$ and $V^{(r+2)}_{\text{par}}$ would be removed, and their weights would be assigned to the past vertical edges. Lastly, we remark that all boundary vertices, including the past and future ancilla vertices, can be merged into a single boundary vertex. In such a setting, each edge of the matching graph $G_r$ encodes both a space-like \textit{and} timelike correction. The timelike component for the edge $e_j$ is obtained by observing if the failure mechanism resulting in the highlighted edge $e_j$ flips the parity of the multi-qubit Pauli measurement. We chose to describe the decoding protocol using \cref{Alg:LatticeSurgery} to avoid figures with multiple edges all incident to the same boundary vertex.

\subsection{Noise model and simulation methodology}

Using the decoding algorithm given in \cref{Alg:LatticeSurgery}, we performed a full circuit-level noise simulation of various code distances and syndrome measurement rounds to estimate the parameters in \cref{eq:Perror} for a $X \otimes X$ measurement. We chose the following biased circuit-level noise model:
\begin{enumerate}
    \item Each single-qubit gate location is followed by a Pauli $Z$ error with probability $\frac{p}{3}$ and Pauli $X$ and $Y$ errors each with probability $\frac{p}{3\eta}$.
	\item Each two-qubit gate is followed by a $\{ Z\otimes I, I \otimes Z, Z \otimes Z \}$ error with probability $p/15$ each, and a $\{X \otimes I, I \otimes X, X \otimes X, Z \otimes X, Y \otimes I, Y \otimes X, I \otimes Y, Y \otimes Z, X \otimes Z, Z \otimes Y, X \otimes Y, Y \otimes Y \}$ each with probability $\frac{p}{15 \eta}$.
	\item With probability $\frac{2p}{3\eta}$, the preparation of the $\ket{0}$ state is replaced by $\ket{1}=X\ket{0}$. Similarly, with probability $\frac{2p}{3}$, the preparation of the $\ket{+}$ state is replaced by $\ket{-}=Z\ket{+}$.
	\item With probability $\frac{2p}{3\eta}$, a single-qubit $Z$ basis measurement outcome is flipped. With probability $\frac{2p}{3}$, a single-qubit $X$-basis measurement outcome is flipped.
	\item Lastly, each idle gate location is followed by a Pauli $Z$ with probability $\frac{p}{3}$, and a $\{X,Y\}$ error each with probability $\frac{p}{3\eta}$.
\end{enumerate}

In our simulations, we chose $\eta = 100$ and for simplicity added a single idle location on the data qubits during the measurement and re-initialization of ancilla qubits. Note that in the limit $\eta \rightarrow 1$, the above noise model reduces to the depolarizing noise model used in Refs.\cite{ChamberlandHeavyHex,ChamberlandColorCode} with the exception that two idle locations were included during measurement and re-initialization of the ancillas. Furthermore, the above noise model assumes that the duration of a CNOT gate is identical to the duration of an ancilla measurement and re-initialization.

\begin{figure}
    \centering
    \includegraphics[width=0.9\columnwidth]{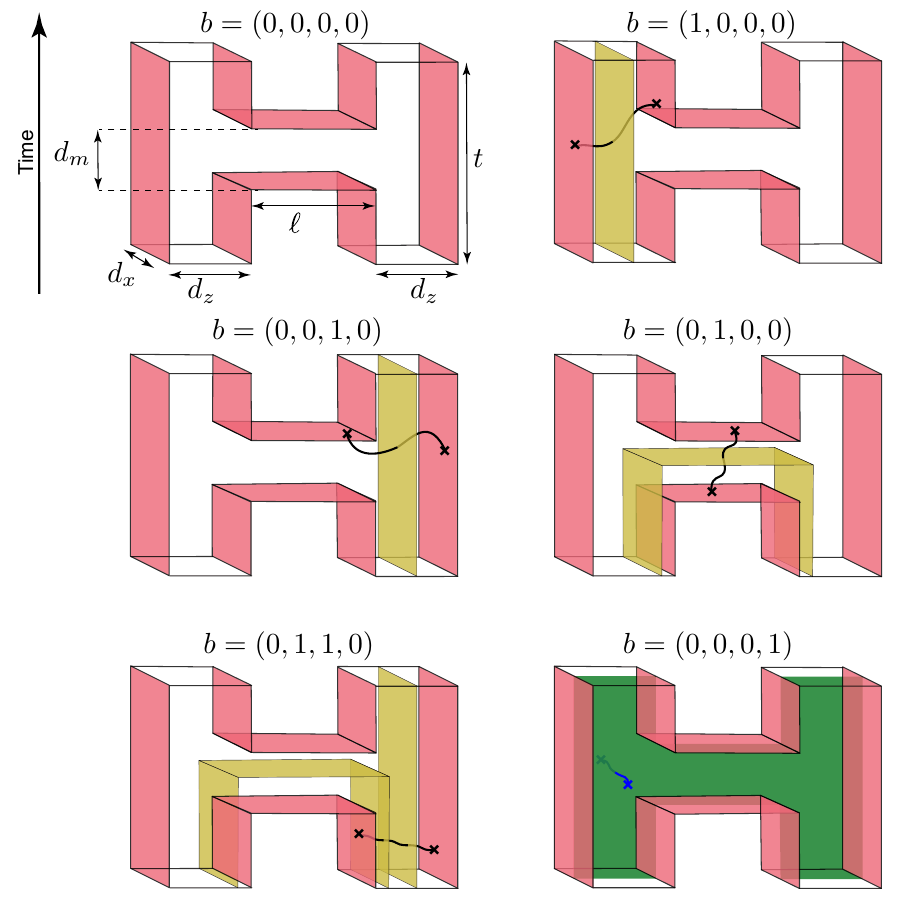}
    \caption{Spacetime diagram of a lattice surgery protocol illustrating different types of failure mechanisms represented by the vector $b=(b_{\text{ZL}},b_{\text{TL}},b_{\text{ZR}},b_{X})$.  For each element $b_j$ in the vector $b$, we associated a logical sheet (shown in yellow or green when triggered) and we have $b_j=1$ if and only if the relevant error type crosses the logical sheet an odd number of times. $Z$-strings trigger yellow logical sheets and terminate on pink boundaries.  $X$-strings trigger green logical sheets and terminate on the foreground and background boundaries (these are transparent for visual clarity). For instance, a $(1,0,0,0)$ event (top right) occurs in the presence of a logical $Z$ string-like excitation on the left surface code patch. Since the excitation does not cross the yellow $\sqcap$ shaped logical sheet, the correct outcome of the multi-qubit Pauli measurement is recorded. A $(0,1,0,0)$ event (middle right) is a pure timelike failure, where the incorrect multi-qubit Pauli measurement outcome is recorded without introducing additional logical $Z$ failures to the two logical patches. This holds because the $Z$ string only crosses the yellow $\sqcap$ shaped logical sheet. A $(0,1,1,0)$ event (bottom left) occurs when a $Z$ string-like excitation crosses the rightmost yellow logical sheet in addition to the yellow $\sqcap$ shaped logical sheet. Such an error results in both a logical $Z$ error on the right logical patch in addition to a timelike lattice surgery failure. Lastly, we illustrate an $X$ string-like excitation crossing the green logical sheet resulting in a logical $X$ failure on both logical patches.}
    \label{fig:FailureMechcanisms}
\end{figure}

For a biased circuit-level noise model (such as the one described above) and an $X \otimes X$ Pauli measurement implemented via lattice surgery, there are fifteen different types of failure mechanisms that can arise during the protocol. We label such failure mechanisms using the binary string $(b_{\text{ZL}},b_{\text{TL}},b_{\text{ZR}},b_{X})$. The bit $b_{\text{ZL}}=1$ corresponds to a logical $Z$ error on the left logical patch, whereas $b_{\text{ZR}}=1$ corresponds to a logical $Z$ error on the right logical patch. The bit $b_{\text{TL}}=1$ indicates a logical timelike failure. Finally, the bit $b_{X}$ indicates if a logical $X$ error occurred during the lattice surgery protocol. Examples of such failure mechanisms using two-dimensional slices of the matching graph are shown in \cref{fig:FailureMechanisms}. For instance, in \cref{fig:MarginalFail}, a series of consecutive measurement errors occur on the same parity vertex, starting in round $r+1$. Since a single measurement error occurs in round $r+1$, the wrong parity of $X \otimes X$ is measured. Due to the series of measurement errors, a single parity vertex near the top boundary is highlighted in $G_r$. The shortest path correction (highlighted in yellow) matches the highlighted parity vertex to the future ancilla vertex. Hence no parity corrections are applied and a logical $(0,1,0,0)$ error occurs. 

Another example is shown in \cref{fig:Fail110}, where a string of $Z$ data-qubit errors results in the highlighted vertices shown in the figure. Prior to the merge, there are no $Z$ errors at the boundary between the logical patches and routing space. Therefore, the correct parity of $X \otimes X$ is measured. However, the minimum-weight path (highlighted in yellow) connecting the highlighted vertex to the future ancilla vertex goes through a transition vertex, so that $v_2 = 1$. The correction thus results in a logical $Z$ error on the left logical patch, in addition to a logical parity measurement failure (since the decoder incorrectly flips the parity) leaving the code with a logical $(1,1,0,0)$ error. Additional examples of failure mechanisms using space-time diagrams instead of matching graphs are provided in \cref{fig:FailureMechcanisms}.

\begin{figure}%[t]
	\includegraphics{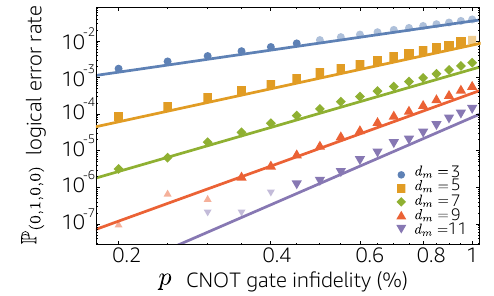}
	\caption{Comparison between the best fit polynomial ${\mathbb{P}}_{(0,1,0,0)}$ for various values of $d_m$ given in \cref{eq:P010} with the data obtained from our Monte Carlo simulations. We chose parameters where $d_x=9$, $d_z=11$, $r=d_z$ and $\ell = 5$. Translucent data points were omitted when obtaining the fitting polynomial in \cref{eq:P010}.}
		\label{Fig:Plotdx9dz11}
\end{figure}

We note that if stabilizer measurements are terminated after $r+d_m$ rounds, higher order failure mechanisms are required to produce logical failures corresponding to $(1,0,0,0)$, $(0,0,1,0)$, $(1,0,1,0)$ and $(1,1,1,0)$ bit strings. Failures corresponding to such strings are thus much less likely compared to $(1,1,0,0)$, $(0,1,0,0)$ and $(0,1,1,0)$\footnote{Note however that if the number of syndrome measurement rounds both before and after the merge are identical, we expect $(1,0,0,0)$ and $(1,1,0,0)$ events to have similar failure probabilities.}. For instance, to obtain a logical failure of the type $(1,0,0,0)$, a logical error on the left logical patch would need to occur without flipping the parity of the $X \otimes X$ measurement. As such, in addition to a logical $Z$ error occurring before the merge, a second failure mechanism would need to occur to undo the wrong parity flip (such as a string of measurement errors like the one shown in \cref{fig:MarginalFail}). Given these observations, we only present the logical failure rate polynomials $\mathbb{P}_{(0,1,0,0)}$, $\mathbb{P}_{(1,1,0,0)}$ and $\mathbb{P}_{(0,1,1,0)}$. Note that due to the high noise bias, we choose a $d_x$ distance such that $\mu \mathbb{P}_{(0,0,0,1)} \le \delta$, where $\mu$ is the number of lattice surgery operations in the algorithm.

Our simulations are performed for syndrome measurement rounds $1$ to $r + d_m$, based on the biased circuit-level noise model described above. The last round is a round of perfect error correction to guarantee projection to the code space. We used \cref{Alg:LatticeSurgery} to correct both spacelike and timelike errors, where each edge in step 7) of the algorithm encodes a particular correction on a subset of the data qubits. Note that we do not perform a round of perfect error correction between rounds $r$ and $r+1$ as was done in Ref.~\cite{Vuillot_2019}. Instead, we performed MWPM using the full syndrome measurement history from rounds 1 to $r+d_m$, and use \cref{Alg:LatticeSurgery} to determine which corrections are applied.

\subsection{Simulation results and conclusions}

Here, we report the outcome of our lattice surgery simulations as summarised by  \cref{eq:P010,eq:P110,eq:P011} and \cref{Fig:Plotdx9dz11}. For each of the dominant failure mechanisms in $(b_{\text{ZL}},b_{\text{TL}},b_{\text{ZR}},b_{X})$, we fit all our data to an ansatz with two free parameters to generate the failure rate polynomials  given by 
\begin{align} \label{eq:P010}
    \mathbb{P}_{(0,1,0,0)} & = 0.01634 d_x  \ell   (21.93 p)^{(d_m+1)/2}, \\
    \label{eq:P110}
    \mathbb{P}_{(1,1,0,0)} & = 0.03148 d_x (28.91 p)^{(d_z+1)/2}, \\
    \label{eq:P011}
    \mathbb{P}_{(0,1,1,0)} & = 0.03 d_x (28.95 p)^{(d_z+1)/2}, \\
    \label{eq:P0001}
    \mathbb{P}_{(0,0,0,1)} & = 0.0148 d_z (0.762 p)^{(d_x+1)/2}.
\end{align}

In \cref{eq:P010}, $\ell$ corresponds to the width of the routing space between the two logical patches. All logical error rate polynomials in \cref{eq:P110,eq:P011,eq:P0001} provide error rates per syndrome measurement round. Per round error rates are computed by varying the number of syndrome measurement rounds for fixed $d_x$ and $d_z$ distances, repeating such procedures for different $d_x$ and $d_z$ distances, and fitting all the obtained data to an ansatz. Note that the $d_z$ distance in \cref{eq:P0001} is taken to be the $d_z$ distance of the full merged surface code patches, whereas $d_z$ in \cref{eq:P110,eq:P011} is the $d_z$ distance of the individual logical patches (since logical $Z$ errors are much less likely to occur when surface code patches are merged due to the increased $d_z$ distance). In \cref{Fig:Plotdx9dz11}, we compare the best-fit polynomial ${\mathbb{P}}_{(0,1,0,0)}$ with a representative subset of our data obtained from our Monte Carlo simulations for various values of $d_m$, where the chosen parameters are described in the caption. The plot shows the exponential suppression in purely timelike error probabilities as a function of $d_m$, and that the data is in good agreement with our best-fit polynomials. In \cref{sec:routing}, after introducing our protocol for minimizing routing costs, we use the logical failure rate polynomials in \cref{eq:P010,eq:P110,eq:P011,eq:P0001} to estimate the overhead costs for implementing quantum algorithms. 

\begin{figure*}
    \centering
    \includegraphics[width=300pt]{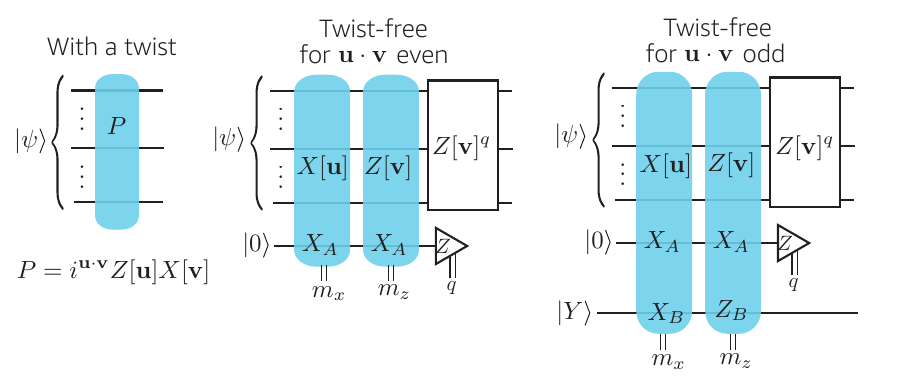}
    \caption{Implementations of a Pauli measurement of $P = i^{\mathbf{u} \cdot \mathbf{v}} X[\mathbf{u}] Z[\mathbf{v}]$.   Whenever $P$ contains any $Y$ terms, simple lattice surgery operations can not be used.  The standard solution is to use twist based lattice surgery.  However, we show the same outcome can be achieved twist-free with an extra $\vert Y \rangle$ ancilla used to handle cases where $X[\mathbf{u}]$ and $Z[\mathbf{v}]$ do not commute.  The twist-free approaches report $m_x \oplus m_z \oplus c$ as the outcome for the measurement of $P$ where $c$ is a constant determined by $\mathbf{u}$ and $\mathbf{v}$.}
    \label{fig:TwistFree}
\end{figure*}

We conclude this section by pointing out that the above labels used in the logical error rate polynomials, which represent different failure mechanisms that can occur during lattice surgery, can be generalized using $k+2$ bits for an arbitrary $P_1 \otimes P_2 \otimes \cdots \otimes P_k$ Pauli measurement. Out of the $k+2$ bits, $k$ bits are used to represent a logical Pauli error on the logical patches which can also flip the parity of the measured Pauli (such as logical $Z$ errors for $X \otimes X$). Another bit represents a logical parity measurement failure. The last bit encodes the logical Pauli error that affects all logical qubits in the merged patch (such as a logical $X$ error during an $X \otimes X$ measurement). 

\section{Protocol for Twist-free lattice surgery}
\label{sec:TwistFreeLattice}

Lattice surgery provides a fault-tolerant way to measure Pauli operators and is well suited for topological codes. However, not all Pauli operators are equally easy to measure. We say an operator is a $XZ$- Pauli when it is a tensor product of $\{ \id, X , Z \}$.  For $XZ$ Pauli operators, standard lattice surgery suffices and a surface code architecture would need only weight 4 stabilizer measurements.  However, for some topological codes such the surface code, measuring Pauli operators containing any $Y$ terms is more difficult.  It has been shown that this can be achieved by introducing a twist defect for each $Y$ in the Pauli operator~\cite{BombinTwists,litinski2018lattice,litinski2019game}. For examples of twist defect lattice surgery, see \cref{fig:Twist} of this work and Fig. 40(d) of Ref.~\cite{litinski2019game}. However, each surface code twist defect requires a stabilizer measurement on 5 physical qubits. This can be very challenging to implement in a 2D architecture with limited connectivity and could require multiple ancilla qubits. The additional ancilla qubits and gates will thus increase the total measurement failure probabilities for weight-five checks. Furthermore, even a single isolated weight-five check will have an impact on the gate scheduling over the whole surface code patch which can introduce additional types of correlated errors.  Lastly, twist-based surface codes coupled with a MWPM decoder have been shown to have a reduced effective code distance \cite{YoderKim17}. As such, we expect twist-based Pauli measurements will suffer a performance loss relative to twist-free Pauli measurements. Any increases in measurement error probabilities during twist-based lattice surgery can be suppressed by extending $d_m$.  In other words, we expect use of twists to increase the runtime of lattice surgery computations. The exact magnitude of this runtime cost is currently unknown and will depend on the precise twist implementation details and the noise model \footnote{An analysis for the impact of twists on the runtime and performance of lattice surgery will appear in future work.}.

Here we outline an alternative twist-free approach to measuring operators containing Pauli $Y$ terms.  The additional cost of supporting Pauli $Y$ terms relative to measuring $XZ$-Pauli operators is roughly a $2 \times$ slowdown in algorithm runtime and a $+2$ additive cost in the number of logical qubits (though we show later that one of these logical qubits can be borrowed from space allocated to routing).  Whether this $2\times$ slowdown is preferable to the slowdown incurred by using twists is an open question, due to a lack of data on twist performance.

\begin{figure*}
    \centering
    \includegraphics[width=2\columnwidth]{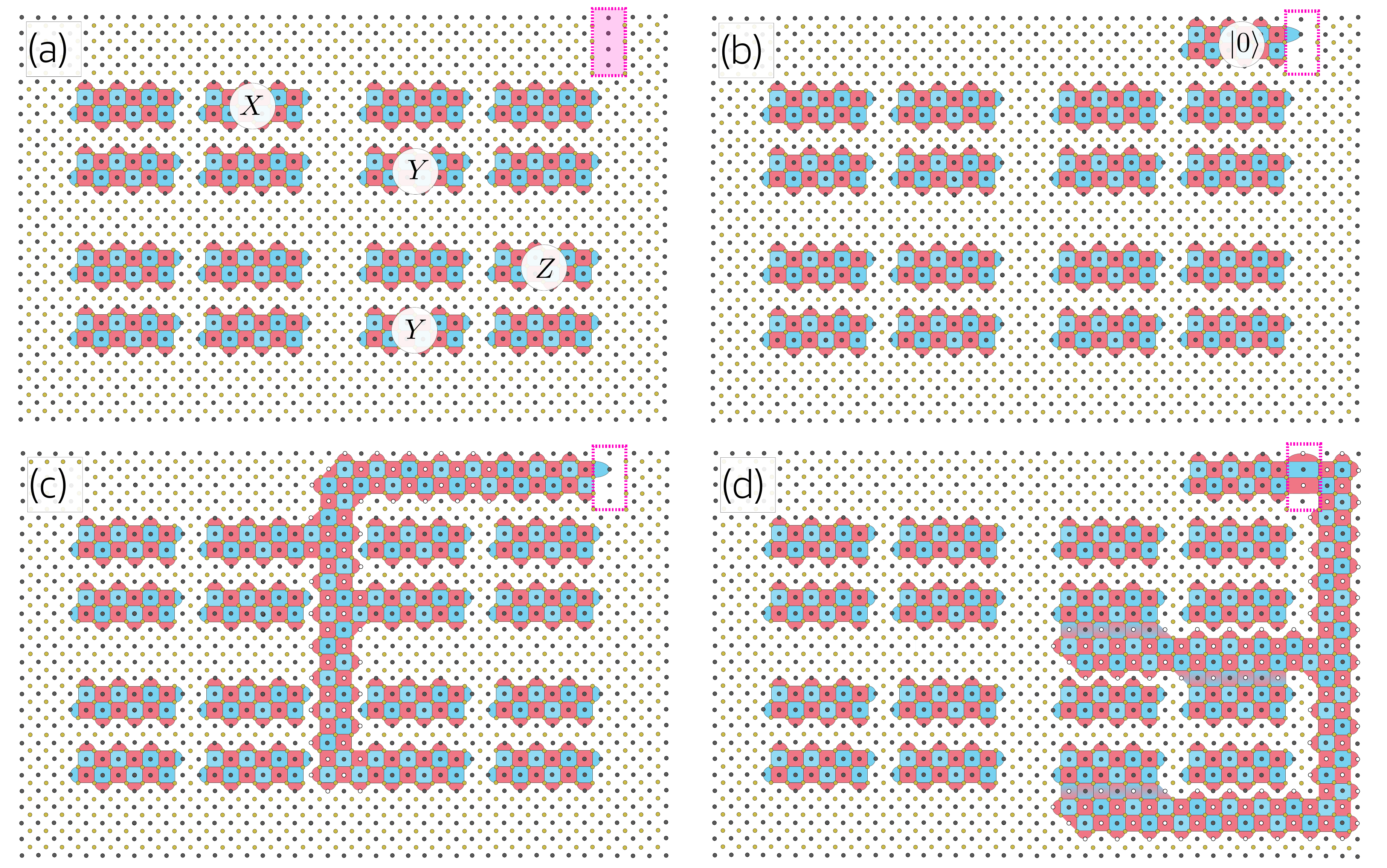}
    \caption{Example of a 2D layout for a quantum computer and the implementation of twist-free lattice surgery operations to realise a $P=Y \otimes Y \otimes X \otimes Z $ Pauli measurement, for which the circuit diagram is given in~\cref{fig:TwistFree}.  In (a), we show an initial layout of thin, rectangular surface codes ($d_x=3$ and $d_z=7$) with labels showing the Pauli operators to be measured. Note the pink rectangle where the hardware layout is slightly adapted to enable elongated stabilizer measurements just within this region.  The space between surface code patches is referred to as \textit{routing space} and will be used in the subsequent steps.  In (b), we show the preparation of a logical $\ket{0}$ state in the routing space. In (c), we show a lattice surgery measurement of $X \otimes X \otimes X \otimes \id \otimes X$.  In (d), we show a lattice surgery measurement of $Z \otimes Z \otimes \id \otimes Z \otimes X$ where at every $Z$ logical boundary we use a domain wall and at the single $X$ logical boundary use elongated stabilizers within the pink region. In both steps (c) and (d), the parity of the logical Pauli measurement is determined by the product of the stabilizers marked with a white vertex, with corrections applied by the decoder.}
    \label{fig:TwistFree2}
\end{figure*}

To explain our protocol, we make use of the notation
\begin{align}
    X[\mathbf{u}] & := \prod_{j=1}^N X^{u_j}_j , \\
    Z[\mathbf{v}] & := \prod_{j=1}^N Z^{v_j}_j ,
\end{align}
for any binary vectors $\mathbf{u}=(u_1, u_2, \ldots u_N)$ and $\mathbf{v}=(v_1, v_2, \ldots v_N)$.  It is well-known that any Hermitian Pauli operator can (up to a $\pm 1$ phase) be decomposed as
\begin{equation}
    P = i^{\mathbf{u} \cdot \mathbf{v}} Z[\mathbf{v}] X[\mathbf{u}]  .
\end{equation}
Then $\mathbf{u} \cdot \mathbf{v} = \sum_j u_j v_j$ counts the number of locations where $X[\mathbf{u}]$ and $Z[\mathbf{v}]$ have overlapping support.  Therefore, using $X_j Z_j \propto Y_j$, we see that $\mathbf{u} \cdot \mathbf{v}$ gives the number of $Y$ terms in $P$. Furthermore, $X[\mathbf{u}]$ and $Z[\mathbf{v}]$ commute whenever $\mathbf{u} \cdot \mathbf{v}$ is even. Equivalently, whenever $P$ contains an even number of $Y$ terms, it can be decomposed (up to a phase) into a product of two commuting operators $X[\mathbf{u}]$ and $Z[\mathbf{v}]$.   

Let assume for now that $\mathbf{u} \cdot \mathbf{v}$ is even, returning to the odd case later. This suggests that we could measure $P$ by using twist-free lattice surgery to measure $X[\mathbf{u}]$ and $Z[\mathbf{v}]$.  However, this would reveal additional unwanted information about $X[\mathbf{u}]$ and $Z[\mathbf{v}]$. To obfuscate this unwanted information, we perform the protocol as illustrated in~\cref{fig:TwistFree} and described below
\begin{enumerate}
    \item Prepare an ancilla (qubit $A$) in the state $\ket{0}$;
    \item Measure $ X[\mathbf{u}] \otimes X_A $ with outcome $m_x \in \{ 0,1 \}$;
    \item Measure $Z[\mathbf{v}] \otimes X_A $ with outcome $m_z \in \{ 0,1 \}$;
    \item Return $m_x \oplus m_z \oplus c$ as outcome of  $P= i^{\mathbf{u}\cdot \mathbf{v} }X[\mathbf{u}] Z[\mathbf{v}]$;
    \item Measure qubit $A$ in the $Z$ basis with outcome $q \in \{ 0,1 \}$;
    \item If $q=1$, then apply a $Z[\mathbf{v}]$  correction (to the Pauli frame).
\end{enumerate}
In step 4, we use a constant $c$ that we define as follows
\begin{align}
    c & = \begin{cases} 0 & \mbox{ if } \mathbf{u \cdot v} = 0 \pmod{4} \\
     1 & \mbox{ if } \mathbf{u \cdot v} = 1 \pmod{4} \\
      1 & \mbox{ if } \mathbf{u \cdot v} = 2 \pmod{4} \\
      0 & \mbox{ if } \mathbf{u \cdot v} = 3 \pmod{4} \\
    \end{cases}
    \end{align}
Clearly, the product of measurement outcomes in steps 2 and 3 gives $X[\mathbf{u}] Z[\mathbf{v}]$ up to some constant.  However, at no point do we learn the value $Z[\mathbf{v}]$ or $X[\mathbf{u}]$, therefore the protocol works as claimed. In \cref{App:TFproof} we provide a more formal proof of the correctness of our protocol and the derivation of the constant $c$. 

We assumed earlier that the Pauli operator $P$ contains an even number of $Y$ terms.  To handle odd numbers of $Y$ terms, we prepare an additional ancilla in the $Y$ basis eigenstate $\ket{Y}=(\ket{0}+i\ket{1})/\sqrt{2}$.  Then, by measuring $Y \otimes P$ we effectively measure $P$.  Furthermore, if $P$ contains an odd number of $Y$ terms, then $Y \otimes P$ contains an even number and can be measured using the above construction.  This modified variant of the twist-free lattice surgery measurement is also illustrated in~\cref{fig:TwistFree}.   Note that the $Y \otimes P$ measurement does not affect the $\ket{Y}$ state and so it can be reused many times and its preparation cost (e.g. through state distillation~\cite{raussendorf2007topological}) only needs to be paid once per algorithm and is therefore negligible.  

% More precisely, we construct a Pauli $\tilde{P}$
% \begin{align}
  %  \tilde{P} & = i^{\tilde{\mathbf{u}} \cdot \tilde{\mathbf{v}}}Z[\tilde{\mathbf{v}}] X[\tilde{\mathbf{u}}]
% \end{align}
% where $\tilde{\mathbf{u}}=(\mathbf{u},1)$ and $\tilde{\mathbf{v}}=(\mathbf{v},1)$.  Then 
% \begin{align}
  %  \tilde{P} & = i^{\mathbf{u} \cdot \mathbf{v} + 1}Z[\mathbf{v}] X[\mathbf{u}] Z_{\ell+1} X_{\ell+1} \\
 %   & = i^{\mathbf{u} \cdot \mathbf{v} + 1}Z[\mathbf{v}] X[\mathbf{u}] (- i  Y_{\ell+1} ) \\
 %     & = i^{\mathbf{u} \cdot \mathbf{v} }Z[\mathbf{v}] X[\mathbf{u}]  Y_{\ell+1} \\
 %      & = P  Y_{\ell+1} .
% \end{align}
% Setting qubit $\ell+1$ to be a $+Y_{\ell+1}$ eigenstate, measuring $\tilde{P}$ will have the same outcome as measuring $P$.  Since $\tilde{\mathbf{u}} \cdot \tilde{\mathbf{v}} = \mathbf{u} \cdot \mathbf{v} + 1$ it follows that we must set $c=0$ when $\mathbf{u} \cdot \mathbf{v}=3 \pmod{4}$ (since then $\tilde{\mathbf{u}} \cdot \tilde{\mathbf{v}}=0 \pmod{4}$)  and  $c=1$ when $\mathbf{u} \cdot \mathbf{v}=1 \pmod{4}$ (since then $\tilde{\mathbf{u}} \cdot \tilde{\mathbf{v}}=2 \pmod{4}$). 

Our twist-free approach uses up to two logical ancillas, a $\ket{0}$ ancilla that is repeatedly reset and sometimes a $\ket{Y}$ ancilla that is reused.  Therefore, we have an additive $+2$ logical qubit cost.  The runtime cost is dominated by steps 2 and 3.  All other steps use only single-qubit operations that effectively take zero time in lattice surgery. Therefore, the runtime has doubled compared to the runtime of measuring a Pauli operator free from $Y$ terms.  If steps 2 and 3 each fail with probability $\mathbb{P}$ (e.g as in \cref{eq:Perror}), then whole protocol fails with probability $\mathbb{P}' =  2 \mathbb{P}(1-\mathbb{P}) \approx 2 \mathbb{P}$. This is a minor effect since failure probabilities are exponentially suppressed by increasing the runtime $d_m$ and using large enough $d_x$ and $d_z$ code distances.

In \cref{fig:TwistFree2}, we show an example of how the circuit picture of Pauli measurements in \cref{fig:TwistFree} can be explicitly mapped into a 2D lattice surgery protocol consisting only of $XZ$-Pauli operator measurements. We present our protocol for thin rectangular surface codes, though our protocol would also work for square surface code patches. First, we notice how the temporary $\ket{0}$ ancilla is prepared in the spare routing space provided for performing lattice surgery, so that it does not actually contribute to the space overhead. Notice also that to accomplish the $ Z[\mathbf{v}] \otimes X_A $ measurement in panel 4, there is one region (highlighted in pink) where we measure elongated stabilizers.  Here we have assumed that the hardware is permanently deformed in this region. In other words, the circuit is \textbf{hardwired} at this location so that the elongated stabilizer can be measured with minimal performance loss compared to any other weight-four stabilizers (for instance, by using longer resonators) and thus does not require additional ancilla qubits. Alternatively, these elongated stabilizers could also be measured in a homogeneous hardware layout but with a modified procedure for performing the measurement. For instance, one could use two ancilla qubits prepared in a GHZ state to measure the elongated stabilizers. However, since the result of the stabilizer measurement would be given by the product of the measurement outcomes of both ancillas, and due to the extra fault locations, using GHZ states would increase the total measurement failure probability of the elongated checks. Another possibility would be to use the second ancilla qubit as a flag qubit \cite{CR1,CR2,CB18,ChamberlandMagic,SCC19,TCL20,chamberland2020very,ChamberlandHeavyHex,ChamberlandColorCode,CR3,ReichardtFlag,TL21}. However, by doing so, one might require an additional time CNOT step per round of stabilizer measurements to perform all two-qubit gates for the stabilizer measurements while avoiding scheduling conflicts.

\section{Temporal encoding for fast lattice surgery}
\label{sec:FastLatticeSurgery}

In \cref{subsec:RevPauliComp,subsec:RevLatticeSurgery}, we discussed the standard approach to Pauli-based computation using lattice surgery and related algorithm runtime bottlenecks. In this section, we show how to exceed this bottleneck and run algorithms at faster clockspeeds using temporal encoding of lattice surgery (TELS).  The key idea behind TELS is to use fast, noisy lattice surgery operations, with this noise corrected by encoding the sequence of Pauli measurements within a classical error correcting code.  This encoding can be thought of as taking place in the time domain, so the encoding does not directly lead to additional qubit overhead costs. Though, there can be a small additive qubit cost when TELS is used for magic state injection, with magic states needing to be stored for slightly longer times.

\subsection{Parallelizable Pauli measurements}
\label{subsec:ParallelPaulis}

Here we review parallelization where the sequence of Pauli measurements can be grouped into  sets of \textit{parallelizable} Pauli measurements. Let $P_{[t, t+k]}:=\{ P_{t}, P_{t+1}, \ldots , P_{t+k}  \}$ be a sub-sequence of our Pauli operators.  We say $P_{[t, t+k]}$ is a parallelizable set if: they all commute; any Clifford corrections can be commuted to the end of the sub-sequence.   For example, we get a parallelizable set whenever we use magic states to perform a $T^{\otimes k}$ gate. We show in \cref{fig:TwoTgates}, several ways to implement $T^{\otimes 2}$ with the PBC approach requiring two parallelizable measurements $\{P_1, P_2 \}=\{C Z_1 Z_3 C^\dagger, C Z_2 Z_4 C^\dagger \}$.   Therefore, given a circuit with $\mu$  $T$-gates and $T$-depth $\gamma$, the Pauli measurement sequence can always be split into a sequence of $\gamma$ parallelizable sets of average size $k:=\mu / \gamma$.
  
Fowler introduced the notion of time-optimal quantum computation~\cite{Fowler2012} and Litinski (see Sec 5.1 of Ref.~\cite{litinski2019game}) showed how this can be realised using lattice surgery in a two-dimensional layout.  In time-optimal PBC, an $n$-qubit computation of $T$-depth $\gamma$ can be reduced to runtime $O(n+\gamma)$.  However, the space-time volume is never compressed by using the time-optimal approach, so that reducing the algorithm runtime to $10\%$ of a seqPBC runtime would require at least a $10\times$ increase in qubit cost. Litinski worked through some highly-paralleziable examples in greater detail, showing: a reduction to 56.5\% of seqPBC runtime would need over $6\times$ the qubit costs; and a reduction to $11\%$ of seqPBC runtime would need over $20\times$ the qubit cost.  The qualifier ``over" in these estimates reflects that an increase in space-time volume will also increase the code distance needed, further increasing the overhead of the time optimal approach by a polylogarthmic factor on top of Litiniski's estimates.  While this is a powerful approach to exploring space-time tradeoffs, early fault-tolerant quantum computers will be qubit limited.  Kim \textit{et al.}~\cite{kim2021fault} also proposed another way to exploit large parallelizable sets, but they used long-range gates not possible in 2D hardware and also made some strong assumptions regarding the speed at which transversal gates can be fault-tolerantly applied.
 
From the above, we see that it is crucial to understand the extent to which algorithms possess potential for parallelization. Fortunately Kim \textit{et al.}~\cite{kim2021fault} have already studied quantum algorithms for chemistry and found $k$ to vary between $9 $ and $14$ depending on the orbitals used, so we will regard this as a practically reasonable range.
  
\begin{figure}
    \centering
    \includegraphics[width= 150pt]{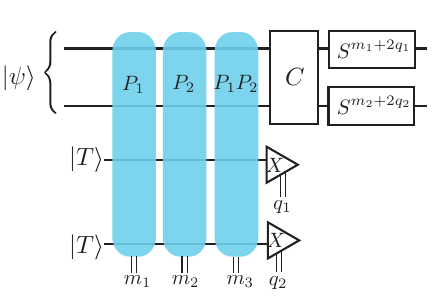}
    \caption{A simple TELS protocol: it is an error detecting version of the PBC used in \cref{fig:TwoTgates} using the measurement code given in \cref{eq:MeasurementCode}. While this approach uses 3 multi-qubit Pauli measurements, the capability to detect an error means that lattice surgery can be executed over a shorter time $d_m$. If there is no error, we will have that the measurement outcomes obey $m_1 \oplus m_2 = m_3$.  When an error is detected we simply remeasure the Pauli operators.}
    \label{fig:TwoTgatesED}
\end{figure}

\subsection{Encodings and code parameter proofs}

Here we introduce our own approach to exploiting parallelizable Pauli sets.  Unlike, previous time-optimal approaches, it does not incur a multiplicative qubit overhead cost and can reduce the overall space-time cost. 

Due to the properties of a parallelizable Pauli set, all Pauli operators within the set can be measured in any order.  Furthermore, we can measure any set $\mathcal{S}$ that generates the group $\langle P_{t}, P_{t+1}, \ldots , P_{t+k} \rangle $.  If the set $\mathcal{S}$ is overcomplete, there will be some linear dependencies between the measurements that can be used to detect (and correct) for any errors in the lattice surgery measurements. For example, consider the simplest parallelizable set $\{ P_1, P_2 \}$ as in \cref{fig:TwoTgates} and let $d_m$ be the required lattice surgery time, so performing both measurements takes $2(d_m+1)$ error correction cycles.  We could instead measure $\{ P_1, P_2, P_1 P_2 \}$. If the 3rd measurement outcome is not equal to the product of the first two measurements, then we know something has gone wrong and can repeat the measurements to gain more certainty of the true values.  By measuring the overcomplete set $\{ P_1, P_2, P_1 P_2 \}$ we have performed an extra lattice surgery measurement, but we have gained that we can tolerate a single lattice surgery failure.  This means that we could instead use $d_m' \ll d_m$ and still achieve the same overall success probability.  If $3 d_m' \ll 2 d_m$ then the computation has been sped-up.  This is the key insight behind temporal encoding of lattice surgery (TELS) and next we dive deeper into more general encoding schemes and their performance. 

In general, given a parallelizable Pauli set 
\begin{align} \label{eq:Unencoded}
    \mathcal{P} = \{ P_{t}, P_{t+1}, \ldots , P_{t+k-1}  \}    
\end{align}
we can define operators generated from this set as follows
\begin{equation}
    Q[\mathbf{x}] := \prod_{j=0}^{k-1}  P_{t+j}^{x_j}
\end{equation}
where $\mathbf{x}$ is a length $k$ binary column vector.  Given a set that generates all the required Pauli operators, so $\langle \mathcal{S} \rangle = \langle \mathcal{P} \rangle$ we can write the elements as 
\begin{align} \label{eq:encoded}
    \mathcal{S} & = \{ Q[\mathbf{x}^{1}], Q[\mathbf{x}^{2}], \ldots , Q[\mathbf{x}^{n}] \}  
\end{align} 
with superscripts denoting different vectors.  Since this is a generating set, the vectors $\{  \mathbf{x}^{1} , \mathbf{x}^{2}, \ldots ,\mathbf{x}^{n} \}$ must span the relevant space.  Furthermore, we can define a matrix $G$ with these vectors as columns and this matrix will specify the TELS protocol.  In the simple $k=2$ example where $\mathcal{S}=\{ P_1, P_2 , P_1 P_2 \}$, we would have that
\begin{equation} \label{eq:MeasurementCode}
    G =\left( \begin{array}{ccc}
       1  & 0 & 1  \\
       0  & 1 & 1  \\
    \end{array}\right) =  \left( \begin{array}{ccc}
       \mathbf{x}^{1}  & \mathbf{x}^{2} &  \mathbf{x}^{3} 
    \end{array}\right)
\end{equation}
Notice that the rows of this matrix generate the codewords of the $[3,2,2]$ classical code.  In general, we will consider $G$ as the generator matrix for the codewords of an $[n,k,d]$ classical code and we call this the \textit{measurement code} for the protocol.  Note that $k$ is the number of (unencoded) Pauli operators in the generating set.  We only consider full-rank $G$ where $k$ equals the number of rows in $G$.  The number $n$ represents how many Pauli measurements we physically perform in the encoded scheme and corresponds to the number of columns in $G$.  The distance $d$ is the lowest weight vector in the row-span of $G$.

Next we show that the code distance $d$ does indeed quantify the ability of TELS to correct errors.  First, we formalise the redundancy in the set of lattice-surgery measurements. For any length $n$ binary vector $\mathbf{u}=(u_1, u_2, \ldots , u_n )$, we have that
\begin{equation}
    \prod_{j : u_j =1} Q[ \mathbf{x}^{j} ] = Q \left[ \sum_{l} u_l \mathbf{x}^{l}  \right]  
\end{equation}
Since the matrix $G$ is full-rank and has more columns than rows, there will exist $\mathbf{u}$ such that $\sum_{j} u_j \mathbf{x}^{j}=0$.  For these $\mathbf{u}$, we have that 
\begin{equation} \label{eq:Redundant}
    \prod_{j : u_j =1} Q[ \mathbf{x}^{j} ] = \id
\end{equation}
Therefore, these $\mathbf{u}$ vectors describe redundancy in the measurements. The condition $\sum_{j} u_j \mathbf{x}^{j}=0$ can be rewritten compactly as $G \mathbf{u} = 0$. Following the convention in coding theory, this set of $\mathbf{u}$ is called the dual of $G$ and denoted
\begin{equation}
    G^\perp := \{ \mathbf{u} :  G \mathbf{u} =0 \pmod{2} \}    
\end{equation}
Next, we consider how this redundancy is used to detect time-like lattice surgery errors. We let $\mathbf{m}=\{ m_1, m_2, \ldots m_n \}$ be a binary vector denoting the outcomes of the lattice surgery Pauli measurements in the set $\mathcal{S}$.  That is, if a measurement of $Q[ \mathbf{x}^{j} ]$ gives outcome ``+1" we set $m_j=0$ and when the measurement of $Q[ \mathbf{x}^{j} ]$ gives ``-1" we set $m_j=1$.  Given a $\mathbf{u} \in G^\perp$, we know the Pauli operators product to the identity (recall \cref{eq:Redundant}) so when there are no time-like lattice surgery errors we have
\begin{equation}
    \prod_{j : u_j =1} m_j = \mathbf{u} \cdot\mathbf{m} = 0 \pmod{2} .
\end{equation}
Conversely, if we observe
\begin{equation}
    \prod_{j : u_j =1} m_j = \mathbf{u} \cdot \mathbf{m} = 1 \pmod{2} 
\end{equation}
then we know a time-like lattice surgery error must have occurred.  Let us write $\mathbf{m}=\mathbf{s} +\mathbf{e} $ where $\mathbf{s}$ is the ideal measurement outcome and $\mathbf{e}$ is the measurement error.   The ideal measurement outcomes are self-consistent and so always satisfy $\mathbf{u} \cdot \mathbf{s} = 0$ for all $\mathbf{u} \in G^\perp$.  Therefore, we see that an error $\mathbf{e}$ is undetected if and only if $\mathbf{u} \cdot \mathbf{e} = 0$ for some $\mathbf{u} \in G^\perp$. This is equivalent to undetected errors $\mathbf{e}$ being in the row-span of $G$ (since the dual of the dual is always the original space).  Recall, the distance $d$ denotes the lowest (non-zero) weight vector in the row-span of $G$.  Therefore, $d$ also denotes the smallest number of time-like lattice surgery errors needed for them to be undetected by TELS.  Consequently, if $\mathbb{P}$ is the probability of a single timelike error, TELS error-detection will fail with probability $O(\mathbb{P}^d)$.

Matrices such as $G$ also appear in the literature in the context of measuring overcomplete sets of stabilizers for some quantum error correction code.  In the error correction setting, these codes have been called measurement codes~\cite{puri2020bias}, meta-checks~\cite{campbell2019theory,quintavalle2021single}, symmetries~\cite{BravyiBiased} and syndrome measurement codes~\cite{ashikhmin2020quantum}.  However, to the best of our knowledge, this is the first time that this idea has been deployed in the context of lattice surgery as a strategy to improve algorithm runtimes. 

\begin{figure*}
    \centering
    \includegraphics{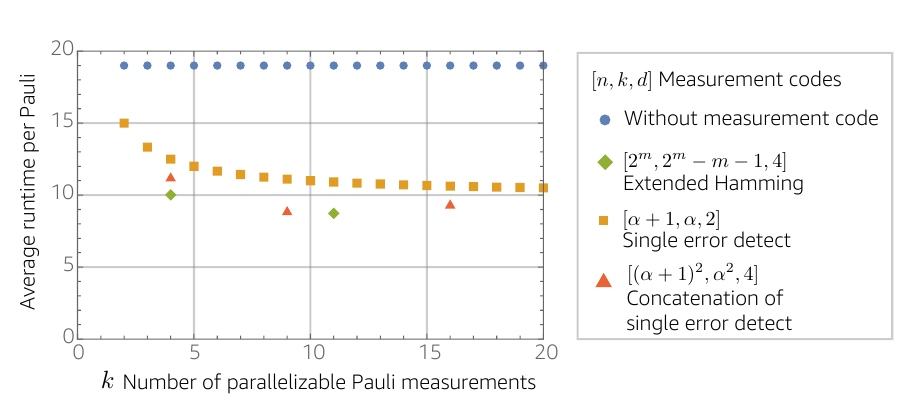}
    \caption{ A comparison of the standard lattice surgery approach (without the measurement code) with 3 different temporally encoded lattice surgery (TELS) schemes for performing a set of $k$ parallelizable Pauli measurements.  We assume that Pauli measurements give an incorrect outcome due to a time-like failure with a probability determined by \cref{eq:P010} with $d_x \ell =100$ and $p=10^{-3}$.  We set the allowed error per Pauli at $\delta=10^{-15}$.  The $[2^m, 2^m - m-1, 4]$ are well-known Hamming codes. The $[\alpha+1,\alpha,2]$ are single error detection codes, and $[(\alpha+1)^2,\alpha^2,4]$ are concatenated single error detection codes. While there are big jumps in $k$ between the best performing codes, these jumps could be partially smoothed out by considering other codes such as concatenated codes with different inner/outer code sizes, such as the $[(\alpha+1)(\beta+1),\alpha \beta ,4]$ codes.   We also considered the triply concatenated codes with parameters $[(\alpha+1)^3,\alpha^3,8]$ but they performed poorly in the parameter regime shown here and so have been omitted for clarity.}
    \label{fig:Fast}
\end{figure*}

\subsection{Examples and numerics}

The simplest examples of TELS will detect a single error.  Given a Pauli set $\{ P_{t}, P_{t+1}, \ldots , P_{t+k}  \}$ we measure each of these observables separately, and then the product of them so that the measurement code has generator matrix
\begin{equation}
    G  = \left( \begin{array}{ccccc}
       1  & 0 & \ldots & 0 & 1  \\
       0  & 1 & \ldots & 0 & 1  \\
       \vdots  & \vdots & \vdots & \vdots & \vdots  \\
       0  & 0 & \ldots & 1 & 1  
    \end{array} \right)
\end{equation}
which is an identity matrix padded with an extra column that is an all 1 vector.  Therefore, this corresponds to a $[\alpha+1,\alpha,2]$ classical code that detects a single error.  Concatenating such a code $m$ times gives a code with parameters $[(\alpha+1)^m,\alpha^m,2^m]$.

We can also consider using a simple $[8,4,4]$ extended Hamming code as the measurement code, with generator matrix
\begin{equation}
    G = \left( \begin{array}{cccccccc}
        0 & 0 & 0 & 0 & 1 & 1 & 1 & 1 \\
        1 & 1 & 1 & 1 & 0 & 0 & 0 & 0 \\
        1 & 1 & 0 & 0 & 1 & 1 & 0 & 0 \\
        1 & 0 & 1 & 0 & 1 & 0 & 1 & 0 \\  
    \end{array}
    \right)
\end{equation}
This corresponds to replacing $\{ P_1, P_2, P_3, P_4 \}$ with $\mathcal{S}$ containing the 8 operators
\begin{align} \label{eq:Sexample}
  \mathcal{S} & = \{  P_2 P_3 P_4 ,  P_2 P_3  , P_2  P_4 ,   P_2  ,  P_1 P_3 P_4 ,  P_1 P_3 , P_1 P_4 ,  P_1 \} 
\end{align}
Because the generator matrix has distance 4, this scheme will detect up to 3 errors.  This Hamming code is the $m=3$ member of a family of $[2^m, 2^m-m-1,4]$ extended Hamming codes.

There are several viable strategies to handle a detected error.  Here we consider the following detect/remeasure strategy: if a distance $d$ measurement code is used with lattice surgery performed for time $d_m$, then whenever an error is detected we ``remeasure" but this time using the original Pauli set $\mathcal{P}$ instead of using the overcomplete set $\mathcal{S}$.  For the remeasure round, we perform lattice surgery using an amount of time $\lceil q d_m \rceil$ where $q$ is some constant scaling factor whose value we discuss shortly.  The expected runtime to execute the protocol is then 
\begin{align} \label{eq:RuntimeED}
  T = n (d_m+1) + p_{\mathrm{d}} k q d_m
\end{align}
where $p_{\mathrm{d}}$ is the probability of detecting an error.  When we do not detect an error, the probability of a failure is $O(p^{d (\frac{d_m+1}{2})}) \approx O(p^{d d_m/2})$.  When we do detect an error, the remeasure round will fail with probability  $O(p^{(q d_m+1)/2}) \approx O(p^{q d_m/2})$.  The total failure probability will then be $O(p^{d d_m/2} + p_{\mathrm{d}} p^{q d_m/2} )$.  Therefore, we can ensure the total failure probability is $O(p^{d d_m/2})$ by setting $q= d$.  However, due to constant factors the optimal choice of $q$ may be slightly different from $q=d$ and so we numerically optimize from this initial guess. When an error detection occurs, this leads to a long delay of time $k q d_m$ to implement the remeasure round, but in practice $p_{\mathrm{d}}$ is so small that this has minimal impact on the expected runtime. 

We could alternatively just measure the overcomplete set $\mathcal{S}$ and run the measurement code in error correction mode with lattice surgery repeated for time $d_m'$.  Then the protocol would fail with probability $\approx O(p^{d d_m'/4})$. Compared to detect/remeasure strategy, we need $d_m' \approx 2 d_m$ to achieve the same failure probability.  The runtime of an error correction scheme will then be
\begin{align} \label{RuntimEC}
  T' = n (d'_m+1) \approx  n (2 d_m+1) .
\end{align}
Compared to \cref{eq:RuntimeED}, in $T'$ we have dropped the second term at the price of roughly doubling the first term.  However, the second term was small because $p_{\mathrm{d}}$ is small, so overall error correction is not favourable compared to our detect/remeasure scheme.

\cref{fig:Fast} shows some example numerical results using distance 2 and 4 codes. For example, when performing $k = 11 $ parallelizable gates the TELS scheme using extended Hamming codes will have a runtime of $46 \%$ that of a standard seqPBC approach that measures the original parallelizable Pauli set $\mathcal{P}$.  Since Kim \textit{et al.}~\cite{kim2021fault} found that interesting quantum algorithms can have average $k$ between 9 and 14, this suggests around a $2.2\times$ speedup due to TELS on practical problems. To obtain a similar speedup, Litinski estimated the time-optimal approach would cost over $6 \times$ in qubit overhead.  The TELS scheme has no multiplicative qubit overhead, though it does have a small additive qubit overhead as all $k$ magic states must be stored for the full duration of the protocol.  However, fault-tolerant algorithms typically have $N \gg 100$ logical qubits, and so the increase in logical qubits $N \rightarrow N+k$ is small.  Indeed, overall the spacetime volume will decrease, which is impossible using Fowler's time optimal approach.

\begin{figure*}[t]
\includegraphics[width=400pt]{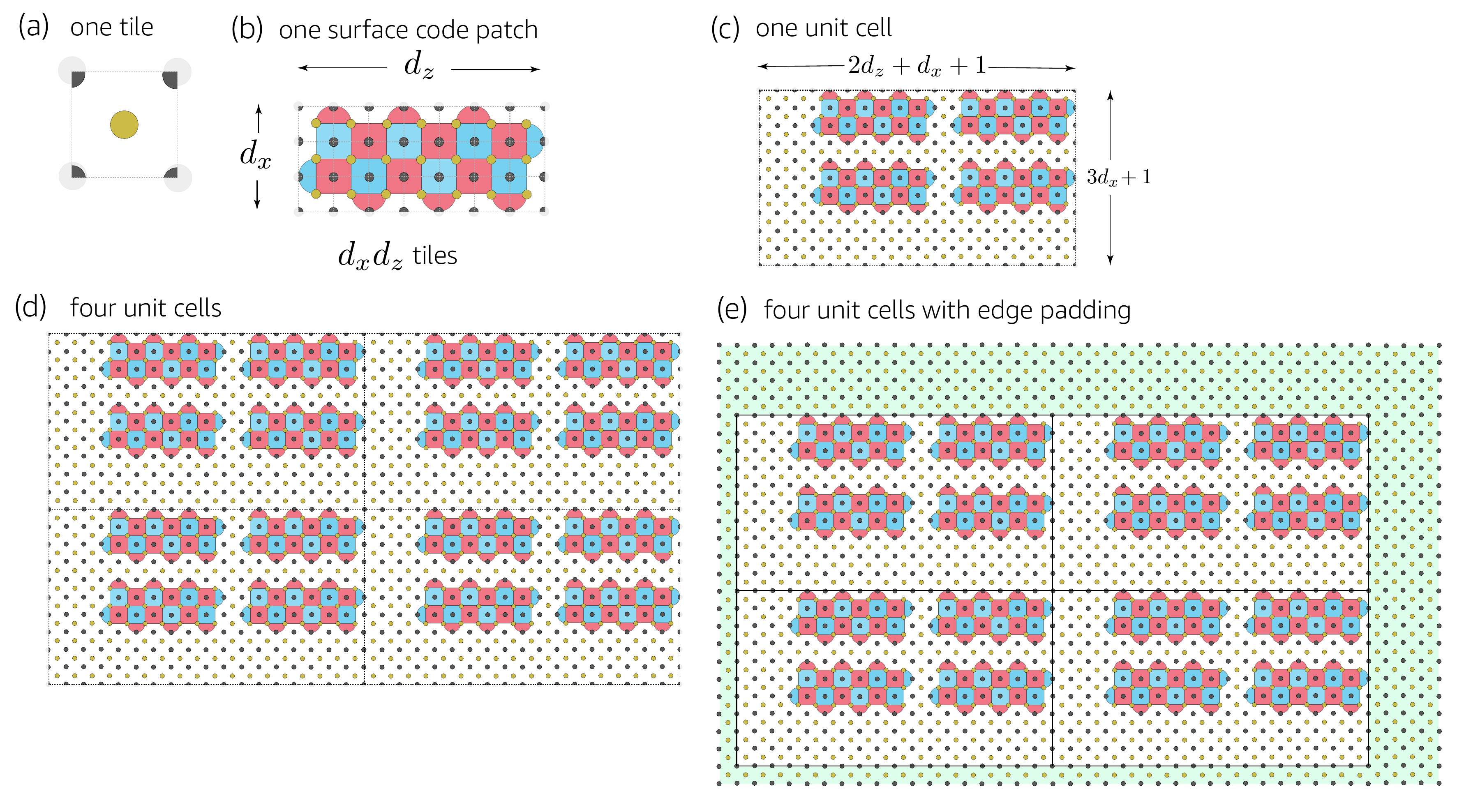}
\caption{Building blocks for a quantum computer. (a) A single tile containing 2 qubits, ancilla qubits (cut at the four corners of the tile) are shared with neighbouring tiles. (b) A single (asymmetric) surface code patch.  The number of tiles needed is the height times width, and so $d_x d_z$. (c) A single unit cell containing four surface code patches and some routing space.  Notice all surface code patches have a $X$ and $Z$ logical boundary adjacent to either routing space within the bulk or the boundary of the unit cell. (d) Four unit cells tiled together. (e) Same as (d) but with the inclusion of extra padding highlighted in green at the edges to provide access to the remaining $X$ and $Z$ logical boundaries and ample routing space. There is more padding on the top and right since we need room the access the remaining $X$ and $Z$ logical operators.}
\label{Fig:UnitCells}
\end{figure*}

We did not find any examples of higher distance codes ($d>4$) that performed better in this parameter regime (e.g. $\delta=10^{-15}$). Going to even lower error rates ($\delta \ll 10^{-15}$) or changing the noise model, then higher distance codes will become useful and the advantage will improve further. Indeed, because of the existence of good classical codes with $n/k=O(1)$ and $d=\Omega(k)$, we know that TELS will asymptotically (for large $k$) be able to execute $k$ parallelizable Pauli measurements in $O(k)$ time and with error $\delta = O(p^{\alpha k})$ for some constant $\alpha$.  In contrast, a standard seqPBC with unencoded lattice surgery would take runtime $O(k \mathrm{polylog}(k))$ to achieve the same error.

\begin{figure}[t]
	\includegraphics[width=\columnwidth]{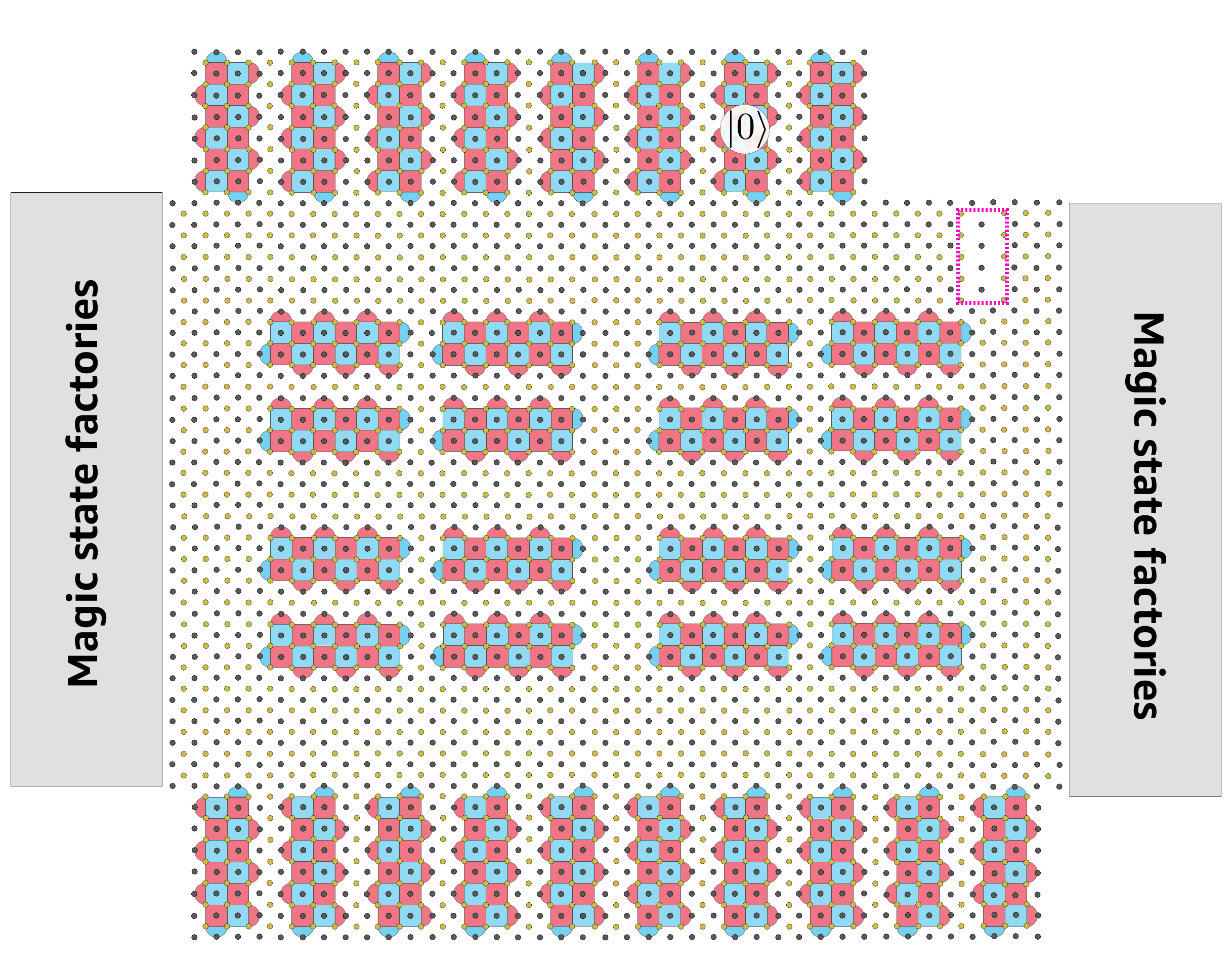}
	\caption{A small example of a quantum computer layout, with a core composed of 4 unit cells (and therefore 16 surface code patches); cache on the top/bottom edge of the core; and magic state factories on the left/right of the core.  A pink, dashed rectangle in the top-right corner indicates a hardwired lattice defect so to enable twist-free lattice surgery.}
		\label{Fig:WholeDevice}
\end{figure}

\section{The core-cache architecture and routing overheads}
\label{sec:routing}

In this section, we discuss a layout and data access structure for a quantum computer that extends on the layout given in panel 1 of \cref{fig:TwistFree2}.  We will consider patches of (possibly rectangular) surface codes of size $d_z$ by $d_x$.   Between these patches we will have some qubits dedicated as a lattice surgery ``bus" or routing space.  We say the routing space supports fast access if logical $X$ and $Z$ operators of every patch are adjacent to the routing space. Litinski proposed several data access structures \cite{litinski2019game}, with his fast data structures using two-tile two-qubit patches (surface code patches that each encode two logical qubits) that are sometimes called hexon surface codes. 

In \cref{Sec:Core} ,  we show that the hexon approach is not necessary, and give a layout for a quantum core (what Litiniski calls a fast data access structure). Furthermore, we show that a lower routing overhead is possible when the surface code patches are thin rectangles (e.g. $d_z \gg d_x$) as is the optimal choice for highly biased noise. In \cref{subsec:cache}, we will also discuss the idea of a core-cache model where logical qubits are temporally moved out of the fast data access structure to reduce the routing overhead.

\subsection{A quantum computer core}
\label{Sec:Core}

\begin{figure*}
    \centering
    \includegraphics{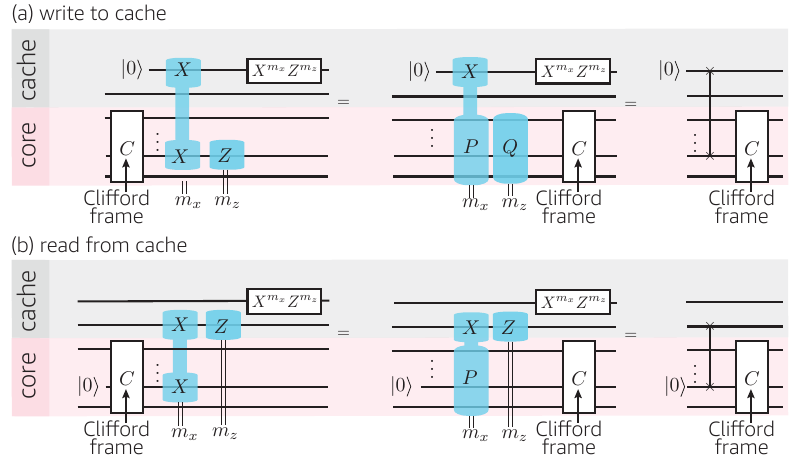}
    \caption{Circuit diagrams for the operations (a) write to cache (WTC); and (b) read from cache (RFC).}
    \label{fig:ReadWriteCache}
\end{figure*}

We count resource costs in terms of square tiles as defined in \cref{Fig:UnitCells}a.  Each tile contains a single data qubit and four quarter ancilla qubits.  Therefore, a device with $T$ tiles will require roughly $2T$ qubits. However, we cannot cut qubits into quarters and so a precise counting will include these.  For instance, a rectangular device with a height of $h$ tiles and width of $w$ tiles would have a total of $T=wh$ tiles and $2T + w + h +1 $ qubits.  When the device is roughly square, then $h$ and $w$ are of size $O(\sqrt{T})$ and so a negligible additive cost compared to $2T$.  We can realise a surface code patch using $d_x d_z$ tiles as in \cref{Fig:UnitCells}b and therefore $2 d_x d_z$ qubits. The number of data and ancilla qubits actively used in the surface code patch is  $2d_xd_z-1$, and so when we try to pack them in a 2D arrangement, the tightest possible packing will contain one idling qubit per patch. 

We collect surface code patches into groups of four, which we call a unit cell (see \cref{Fig:UnitCells}c). These unit cells are then repeated as shown in \cref{Fig:UnitCells}d to get the required number of logical qubits. Furthermore, we arrange the unit cells to form a quantum ``core", and assume some additional padding shown in \cref{Fig:UnitCells}e.  Notice that in \cref{Fig:UnitCells}, every patch has logical $X$ and $Z$ boundary operators connected to the routing space, which enables us to quickly perform multi-qubit Pauli measurements between any subset of qubits within the core. Additionally, there are unused qubits between some of the surface code patches.  The spacing of the qubits ensures that lattice surgery can be performed (as we saw in \cref{fig:TwistFree2}) without using lattice twists that incur additional practical difficulties to implement in fixed and low connectivity hardware.  In contrast, the data access structures proposed by Litinski~\cite{litinski2019game} assumed liberal use of twists.

The routing overhead for unit cells is then the ratio of the number of tiles divided by the cost without any routing space (e.g. $4 d_z d_x$).
\begin{align}
	O^{\textrm{(unit cell)}}_{(d_z, d_x)} & = \frac{(2 d_z + d_x + 1)(3d_x+1) }{4 d_z d_x}. 
\end{align}
The overhead for the entire core will include a contribution from the additional padding shown in \cref{Fig:UnitCells}e.  However, in the limit of many unit cells, the total overhead is dominated by the unit cell overhead. In the limit of large distances $d_z, d_x \gg 1$, we have
\begin{align} \label{OverheadUnit}
	O^{\textrm{(unit cell)}}_{(d_z, d_x)}  & \approx \frac{3}{2} +  \frac{3}{4}\frac{d_x}{ d_z}.
\end{align}
Therefore, in the limit of large noise bias, $d_z \gg d_x$, the routing overhead factor is $1.5 \times$.  We can compare this with the fast data blocks  of Litinski (see e.g. Figure 23 of  \cite{litinski2019game} ) where the overhead factor is $2 \times$ (in the large device limit) and so is more expensive.  In contrast,  for unbiased noise and $d_z= d_x$ our scheme has an asymptotic routing overhead factor of $9/4$, which is slightly worse than Litinski's $2\times$ routing overhead.  Indeed, solving \cref{OverheadUnit} equal to 2 shows that $d_x < (2/3) d_z$ is the condition for our approach to have a routing overhead advantage. A more general analysis of the routing overhead which includes the green padded regions shown in \cref{Fig:UnitCells}e and contributions from the cache (see \cref{subsec:cache}) is given in \cref{App:OverheadCalcs}. We also point out that routing overhead is not the only important figure of merit. Our proposed design avoids twist defects and other significant lattice irregularities, and therefore may be useful even without noise bias.

\begin{figure*}
    \centering
    \includegraphics[width=430pt]{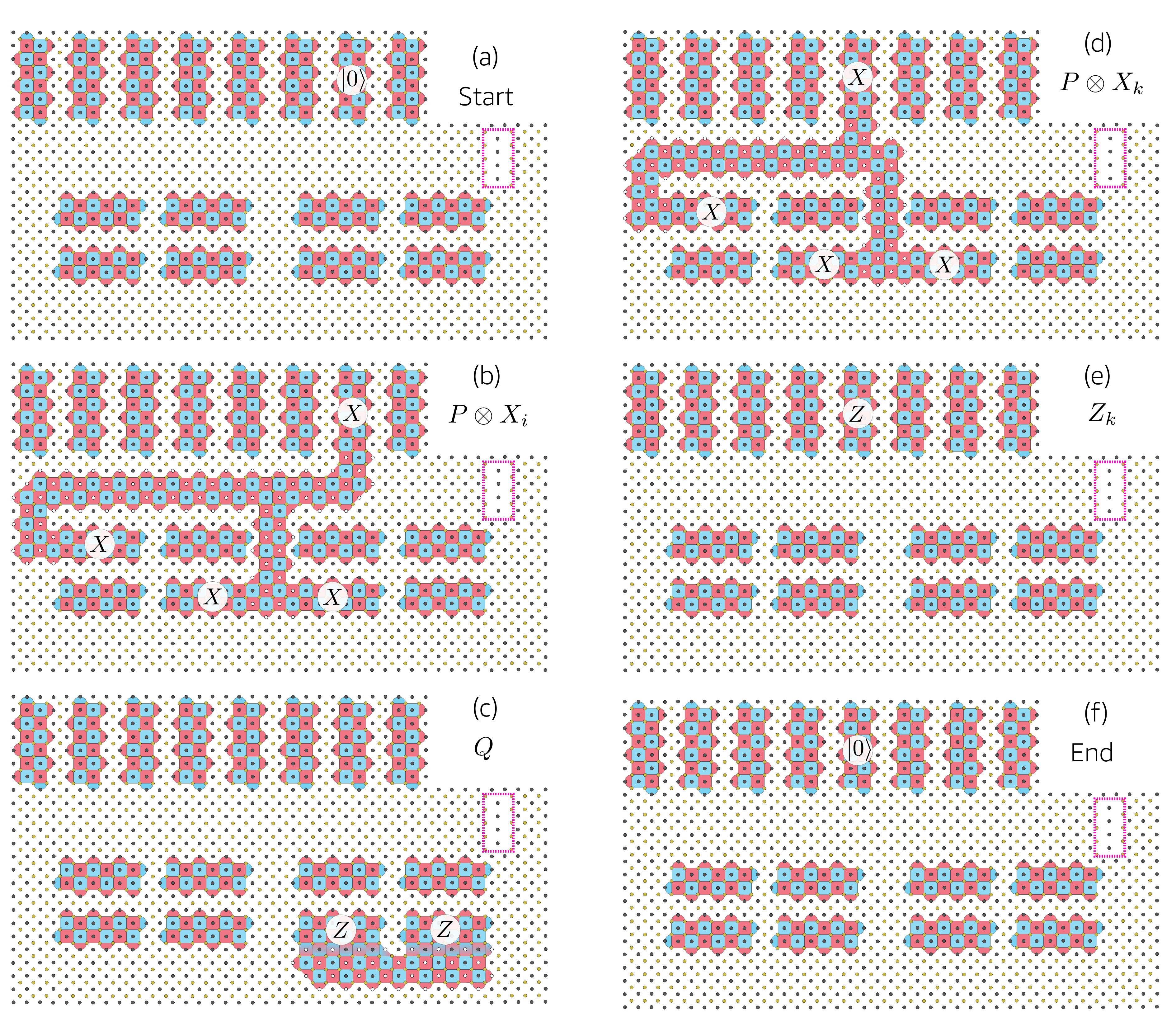}
    \caption{Example lattice surgery diagrams for: using write to cache and read from cache to exchange qubits $j$ and $k$ between core and cache. (a) Illustration of the initial configuration with a $\vert 0 \rangle$ ancilla (index label $i$) in the cache. Qubit $j$ is in the core and qubit $k$ is in the cache and we wish to swap their locations. (b)  We measure $P \otimes X_i$ where $P=C X_{j} C^\dagger$ where $C$ is the Clifford frame.  For simplicity, we assume $P$ is composed of only $X$ operators.  (c) We measure $Q= C Z_j C^\dagger$ and this time assume it is composed of only $Z$ operators. (d) We measure $P \otimes X_k$ where $P=C X_{j} C^\dagger$ is the same operator as in (b). (e) We measure the single qubit Pauli $Z_k$.  If either of the simplifying assumptions we made for $P$ or $Q$ do not hold, then we will need to use the twist-free protocol (or use twists). (f) At the end of the protocol, qubit $j$ is in the cache, qubit $k$ is in the core, and there is a \ket{0} ancilla in the cache ready for further swaps.}
    \label{fig:PushPull}
\end{figure*}

\subsection{A quantum computer cache}
\label{subsec:cache}

We now proceed to build additional structure around the core.  Using state distillation to prepare magic states, we will need factories that supply the core.  The purely fast data access approach prioritises speed over qubit cost.  Here we also discuss the idea of a core and cache architecture, where some logical qubits are temporarily stored in a quantum analog of cache.  However, with some time cost, logical qubits can be quickly swapped in and out of the cache.  A small-scale sketch of a device comprising core, cache and magic state factories is illustrated in~\cref{Fig:WholeDevice}.  

Packing qubits more compactly in the cache will clearly reduce the overhead costs.  However, such a layout comes at a price since only the $X$ logical operators of these qubits can be accessed when it is in the cache. To access the logical $Z$ operators, it must be swapped out of the cache and into the core. For a qubit stored in the cache, we can perform the following operations
\begin{enumerate}
      \item Perform single qubit $X$ or $Z$ measurements for a qubit in the cache (time cost: zero).
      \item Measure multi-qubit operators of the form $A \otimes B$ where $B$ acts on the cache qubits and is a tensor product of $X$ operators only, and $A$ acts on the core qubits and can be an arbitary Pauli operator. (time cost: $d_m$).
      \item Perform Pauli updates to the Pauli frame (in software).
    \item Perform Cliffords to qubits in the core, by updating the Clifford frame (in software).
 \end{enumerate}
For algorithms where swapping in/out of the cache can be made infrequently (compared to other time costs), our approach will reduce the routing overhead with a mild impact on the algorithm runtime. Note also that our core-cache architecture can be used in combination with the twist-free or TELS schemes already proposed.

This leaves the question of how to swap the location of qubits from the cache to the core.  We cannot directly implement the Clifford SWAP operation, since Clifford operations can only be performed on core qubits. Furthermore, surface code patches cannot be moved around to swap their positions since such operations would require the Clifford frame $C$ to be relabelled. Such a relabelling might make $C$ act non-trivially on qubits in the cache. Rather, when performing a SWAP from a qubit in the core to the cache, we need to clean the Clifford frame so it only acts on core qubits. 

We now define two elementary operations: write to cache (WTC) and read from cache (RFC), which when combined will enable a Clifford-cleaned swap.  We first present the WTC and RFC protocols using circuit diagrams in \cref{fig:ReadWriteCache}. They are both essentially 2-qubit teleportation protocols, but with the Clifford frame adapting the Pauli measurements performed.  The WTC operation uses two multi-qubit Pauli measurements, whereas RFC can be performed faster as it uses only one multi-qubit Pauli measurement and a single-qubit Pauli measurement (that takes zero time). The WTC operation requires a logical $\ket{0}$ ancilla in the cache, which through the protocol swaps place with a logical qubit in the core.  The RFC operation requires a logical $\ket{0}$ ancilla in the core, which through the protocol swaps place with a logical qubit in the cache.

For a pair of logical qubits, one in the core and one in the cache, we can cleanly SWAP their positions, by executing WTC followed by RFC. The $\ket{0}$ ancilla initially in the cache, will move to the core, then back to the cache, but to a different cache location than it started.  The whole swap procedure requires three multi-qubit Pauli measurements. \cref{fig:PushPull} shows an example using lattice surgery.   This figure shows a simple scenario where these three multi-qubit Pauli measurements are $XZ$ Pauli measurements (even after conjugated by the Clifford frame) and so can be realised with three simple lattice surgery operations.  However, more generally when some measurements are not of $XZ$-type, we need to either: use twist defects (and benchmark their performance); or use our twist-free protocol and realise the swap with up to six lattice surgery operations.

In \cref{App:OverheadCalcs}, we perform a resource cost analysis for simulating the Hubbard model using the full core-cache architecture described in this section. In particular, we provide a rigorous analysis of routing overhead costs including contributions from the cache and green padded region in \cref{Fig:UnitCells}e.

\section{Conclusion}
\label{sec:Conclusion}

In \cref{sec:LatticeSurgeryDecoder} we introduced a decoding algorithm compatible with lattice surgery protocols and numerically computed failure rate polynomials for the dominant failure mechanisms of an $X \otimes X$ Pauli measurement. Our analysis allows one to compute appropriate $d_x$ and $d_z$ code distances, as well as the number of syndrome measurement rounds $d_m$ during lattice surgery for successfully implementing algorithms. 

In \cref{sec:TwistFreeLattice}, we introduced a twist free protocol for measuring arbitrary Pauli operators using lattice surgery. The protocol incurs a $2 \times$ slowdown in algorithm runtime. However surface codes with twists require higher weight stabilizer measurements, increased gate scheduling complexities and have a reduced effective code distance when using a MWPM decoder. Such features will inevitably cause a reduction in performance compared to lattice surgery protocols involving only $X$ and $Z$ Pauli measurements. Consequently, a careful numerical analysis with twists is needed in order to determine whether using twists can beat the $2\times$ cost of the twist-free approach.

In \cref{sec:FastLatticeSurgery} we introduced a technique which we call temporal encoding of lattice surgery. By encoding lattice surgery measurements in the time domain, we showed that a $2\times$ reduction in algorithm runtimes can be achieved for quantum algorithms of practical scale without incurring additional qubit overhead costs. For more highly parallelizable algorithms or larger algorithms, using larger classical code distances and exploring other code families will lead to even greater improvements in algorithm runtimes. After posting a pre-print of this work, Craig Gidney emailed us and later posted on Twitter, remarking that temporal encoding of lattice surgery could be used to reduce the runtime of magic state distillation factories.

Lastly, in \cref{sec:routing}, we provided a core-cache architecture compatible with our new lattice surgery protocols. A subset of the data qubits are stored in a cache, which reduces the footprint of the routing space, and can be quickly accessed when $Z$ measurements need to be performed. We found that for such an architecture, routing overhead costs add a factor of $1.5 \times$ to the total resource costs for performing lattice surgery. A clear direction for future work would be to analyze such architectures in the presence of twists in order to better understand the tradeoffs with using a twist-free lattice surgery protocol.

Apart from considering twists, a direction of future work would be to apply our methods using other error correcting codes to potentially achieve lower resource costs. Promising code candidates include codes tailored for biased noise such as the XZZX surface codes~\cite{TuckettBiasedNoise,BravyiBiased,XZZXcodes,PuriXZZX}, subsystem codes with high thresholds~\cite{HiggottSubsystem} and other codes families with high encoding rates such as hyperbolic surface codes~\cite{Breuckmann_2017,Conrad2017}.

\section{Acknowledgments}
\label{sec:Acknowledgments}
We thank Giacomo Torlai for his help with using the AWS clusters. We thank Oscar Higgott, Fernando Brandao and Noah Shutty for their comments and suggestions on our manuscript.  

\appendix

\section{Twist-free proof}
\label{App:TFproof}

Here we give a formal proof that the twist-free lattice surgery protocol works as claimed. Consider the case when step 5 yields a $q=0$ outcome so we project onto the $\ket{0}$ state.  Then steps 1-5 implement
\begin{align}
  M_{+1}  & : =   \kb{0}{0}_A \Pi_{ZX}(m_z) \Pi_{XX}(m_x)  \kb{0}{0}_A , 
\end{align}
where
\begin{align}
    \Pi_{ZX}(m_z) & :=\frac{1}{2} \left( \id \otimes \id + (-1)^{m_z}  Z[\mathbf{v}] \otimes X_A  \right) \\
    \Pi_{XX}(m_x) & := \frac{1}{2} \left( \id \otimes \id  + (-1)^{m_x}  X[\mathbf{u}] \otimes X_A    \right) 
\end{align}
Using that for arbitrary $Q$ 
\begin{align}
 \kb{0}{0}_A ( Q \otimes \id  ) \kb{0}{0}_A  & = Q \otimes \kb{0}{0}_A  , \\ \nonumber
 \kb{0}{0}_A (  Q \otimes X_A   ) \kb{0}{0}_A  & =  0 ,
\end{align}
we deduce
\begin{equation}
  M_{+1} = \frac{1}{4}    \left[  \id +  (-1)^{m_x+m_z}  (Z[\mathbf{v}] X[\mathbf{u}]) \right] \otimes  \kb{0}{0}_A ,
\end{equation}
which is proportional to the projector for a $Z[\mathbf{v}] X[\mathbf{u}]$ measurement with outcome $m_x \oplus m_z$.  When $\mathbf{u}\cdot \mathbf{v} = 0 \pmod{4}$, we have $P=Z[\mathbf{v}] X[\mathbf{u}]$ and so  $m_x \oplus m_z$ is the outcome of measuring $P$  (justifying $c=0$ in this case).  On the other hand, if $\mathbf{u}\cdot \mathbf{v} = 2 \pmod{4}$, we have $P=-Z[\mathbf{v}] X[\mathbf{u}]$ and so  $m_x \oplus m_z \oplus 1$ is the outcome of measuring $P$  (justifying $c=1$ in this case).  Recall that we are currently assuming $\mathbf{u}\cdot \mathbf{v}$ is even.

In the event that step 5 yields a $q=1$ outcome, we have that 
\begin{align}
  M_{-1}  & : =   \kb{1}{1}_A \Pi_{ZX}(m_z) \Pi_{XX}(m_x)  \kb{0}{0}_A , 
\end{align}
Using that for arbitrary $Q$ 
\begin{align}
 \kb{1}{1}(Q \otimes \id)  \kb{0}{0}  & =  0 , \\ \nonumber
 \kb{1}{1}  ( Q \otimes X_A)  \kb{0}{0}  & =    Q \otimes \kb{1}{0} .
\end{align}
We have
\begin{align}
  M_{-1} & = \frac{1}{4}  \left[ (-1)^{m_z} Z[\mathbf{v}]    + (-1)^{m_x} X[\mathbf{u}] \right]  \otimes \kb{1}{0}_A  , \\
  & = ((-1)^{m_z} Z[\mathbf{v}]) \otimes X_A M_{+1} .
\end{align}
Therefore, we see that $M_{-1}$ differs from $M_{+1}$ by a $Z[\mathbf{v}]$ correction that we perform in step 6.  There is also a global phase $(-1)^{m_z}$ but this is unimportant.

\CC{We remark that for the case where $\mathbf{u \cdot v} = 1 \pmod{4}$, adding an additional $Y$ operator to $P$ and performing the measurement using the $\ket{Y}$ ancilla is identical to the case where $\mathbf{u \cdot v} = 2 \pmod{4}$. Hence $c$ will be 1. A similar argument can be used to show that $c=0$ for the case where $\mathbf{u \cdot v} = 3 \pmod{4}$.}

\section{Treating space-time correlated edges incident to parity vertices}
\label{App:SpaceTime}

\begin{figure}%[t]
	\centering
	\subfloat[\label{fig:LatticeScheduling}]{%
		\includegraphics[width=0.48\textwidth]{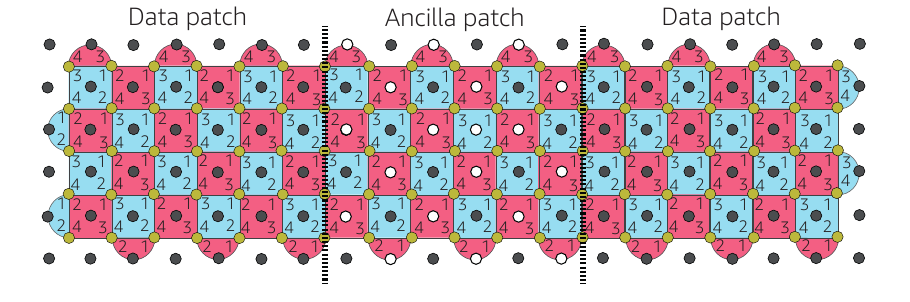}
	}
	\vfill
	\subfloat[\label{fig:SpaceTime1}]{%
		\includegraphics[width=0.48\textwidth]{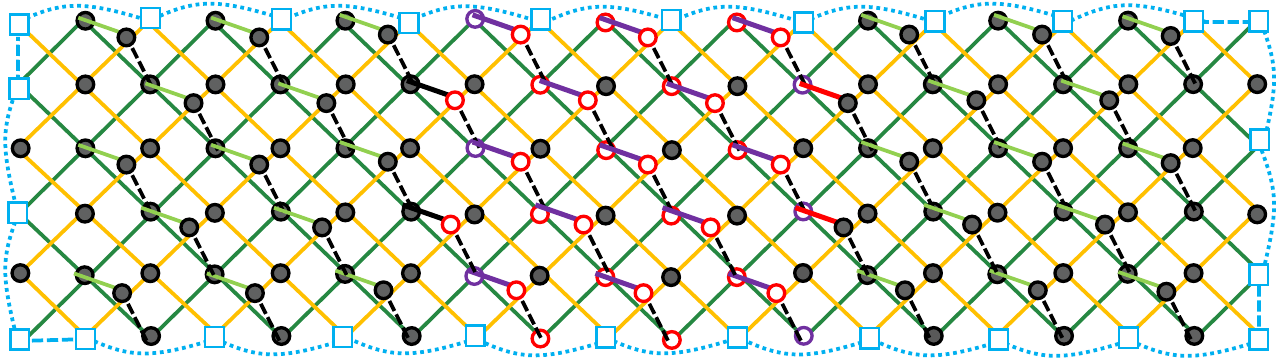}
	}
	\vfill
	\subfloat[\label{fig:SpaceTime2}]{%
		\includegraphics[width=0.48\textwidth]{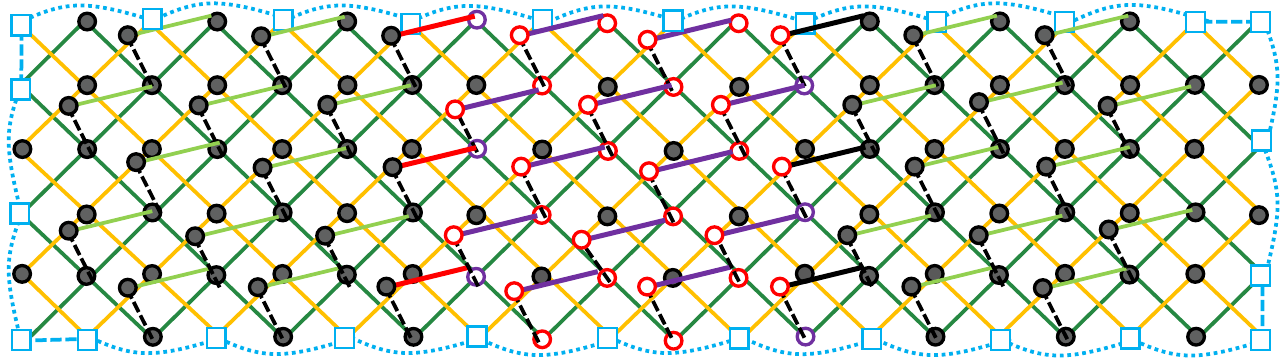}
	}
	\caption{(a) Surface code lattice with $d_x=5$, $d_z=7$. The numbers in each check indicate the gate scheduling of the two-qubit gates (which in our simulations are CNOT gates). (b) First type of space-time correlated edges for $X$-stabilizers. (c) Second type of space-time correlated edges for $X$-stabilizers. Space-time correlated egdes incident to vertices present in the routing space are assigned different colors than green for the reasons described in the text.}
	\label{fig:SpaceTimeEdges}
\end{figure}

As mentioned in \cref{sec:LatticeSurgeryDecoder}, the space-time correlated edges incident to the parity vertices during the first round of the merged surface code patches (i.e. vertices $v \in V^{(r+1)}_{\text{par}}$) need to be treated with care. Such edges are highlighted in red, black and purple in \cref{fig:SpaceTime1,fig:SpaceTime2} for an $X \otimes X$ measurement. 

First, in round $r+1$, edges highlighted in purple must have infinite weight and can thus be removed. This is due to the fact that when the logical patches are merged using the routing space in round $r+1$, individual $X$-stabilizer measurements performed in the routing space (whose ancilla's are marked by white vertices) will have random outcomes and thus cannot be highlighted. As such, failures arising from two-qubit gates performed in the routing space, and which introduce errors that do not interact with stabilizers belonging to logical patches, cannot generate a non-trivial measurement outcome between two consecutive rounds of stabilizer measurements. Although the purple edges discussed above must be removed when they are incident to vertices in rounds $r+1$ and $r+2$, such edges incident to vertices $v_i$ and $v_j$ belonging rounds greater than $r+1$ must be included since they will have finite weights.

Second, space-time edges incident to a single parity vertex $v \in V^{(r+1)}_{\text{par}}$ are highlighted in red in \cref{fig:SpaceTime1,fig:SpaceTime2}. Note that in round $r+1$, such edges are incident to parity vertices colored in purple, which are also transition vertices. Such edges arise from two-qubit gate failures in round $j \ge r+1$ with the property that the errors introduced by the failure only flip the parity vertices in that round. In round $j+1$, the error is detected by $X$-stabilizers belonging to both logical patches and the routing region. Errors introduced by such failures can flip the parity of the $X \otimes X$ measurement outcome. As such, $v_2$ (defined in \cref{Alg:LatticeSurgery}) should also include the number of highlighted red space-time correlated edges incident to transition vertices.

Third, space-time correlated edges highlighted in black have the same effect for correcting errors as all other space-time correlated edges belonging to the logical patches (highlighted in green). The reason is that they are incident to parity vertices in rounds $j \ge r + 2$ and thus failure mechanisms leading to such edges cannot flip the parity of the $X \otimes X$ measurement outcome. 

Lastly, we note that a two-qubit gate failure arising in round $r+1$ and which results in a red highlighted space-time correlated edge will only highlight a single vertex (belonging to the logical patch in round $r+2$) throughout the entire syndrome measurement history (assuming no other failures occur). This is due to the random outcomes of $X$-stabilizers in round $r+1$ (so that vertices for such stabilizers cannot be highlighted in round $r+1$). Since there is an asymmetry between the number of red space-time correlated edges incident to the left data qubit patch and those incident to the right data qubit patch, an asymmetry in the logical failure rate polynomials $\mathbb{P}_{(1,0,0,0)}$ and $\mathbb{P}_{(0,0,1,0)}$ (defined in \cref{sec:LatticeSurgeryDecoder}) will also arise.

\section{Resource cost analysis of the Hubbard model}
\label{App:OverheadCalcs}

\begin{table*}
\begin{tabular}{cccccccccc}  \toprule
size &  &  &  &  &  & $\#$ of physical & $\#$ of logical & $\#$ of logical & $O^{(\text{total})}_{(d_x,d_z,h,w)}$\\
 $L$ & $h$ & $w$ & $d_x$ & $d_z$ & $d_m$ & qubits & qubits (core) & qubits (cache) &  \\ \midrule
 \multicolumn{10}{c}{$u/ \tau=8$} \\ \midrule
 8 & 2 & 6 & 7 & 13 & 12 & 46472 & 48 & 115 & 1.57 \\
 8 & 6 & 6 & 7 & 13 & 12 & 63992 & 144 & 19 & 2.16  \\
 32 & 6 & 8 & 7 & 15 & 12 & 657276 & 192 & 2371 & 1.23  \\
 32 & 14 & 18 & 7 & 15 & 12 & 812532 & 1008 & 1555 & 1.51  \\
  \bottomrule
\end{tabular}
\caption{For a Hubbard model simulation with lattice size $L$ and unit cells of height $h$ and width $w$ in the core, we provide the minimum values of $d_x$, $d_z$ and $d_m$ such that \cref{eq:MarTot,eq:ZLTot,eq:XLTot} are satisfied with $\delta \sim 1 \%$. For the given parameters, we also provide the total number of physical qubits using \cref{eq:Nphys}, and the number of logical qubits in the core and in the cache. The last column includes the multiplicative factor (defined in \cref{eq:Ototal}) that is added to the physical qubit overhead which takes into account routing costs. The values for $N_{\text{TOF}}$ and $N_{T}$ used in computing $\mu$ are obtained from Ref.\cite{CampbellHubbard}. All resource costs exclude contributions from the magic state factory.}
    \label{tab:HubbardModelOverhead}
\end{table*}

Following \cite{CampbellHubbard}, the total number of logical qubits used for simulating a Hubbard model of lattice size $L$ is $N_Q = 2L^2 + \frac{L^2}{2} + 2$. If $T$ gates are performed by catalysis, then an extra logical qubit is needed. We also add another logical qubit for the logical $\ket{0}$ required in the WTC/RFC protocol described in \cref{subsec:cache}.  Using the core-cache model shown in \cref{Fig:WholeDevice}, let $N_1$ be the number of logical qubits in the core, and $N_2$ the number of logical qubits in the cache. Adding the cost of the green padding shown in \cref{Fig:UnitCells}e, we now compute the total routing overhead cost in the core with $h$ unit cells stacked in the vertical direction, and $w$ unit cells stacked in the horizontal direction ($h=w=2$ in \cref{Fig:WholeDevice}d). We begin by defining the following functions which count the number of tiles in the green padded region of \cref{Fig:UnitCells}e, where we have separated the region into four sections
\begin{align} \nonumber
    s_1(d_x,h) & = h(3d_x+1), \\ \nonumber
    s_2(d_x,d_z,w) & = d_x + 2 + w (2d_z+d_x+1), \\ \nonumber
    s_3(d_x,h) & = h(3d_x+1)(d_x+1), \\ \nonumber
    s_4(d_x,d_z,w) & = (d_x+1) \Big( w(2d_z+d_x+1) + d_x + 2 \Big )\ .
\end{align}
The total number of tiles in the green padded region is then given by
\begin{align}
S_{\text{TRGP}}(d_x,d_z,h,w) = & s_1(d_x,h) + s_2(d_x,d_z,w) \nonumber \\ 
& + s_3(d_x,h) + s_4(d_x,d_z,w).
\end{align}
The total routing overhead in the core is then
\begin{align}
    O^{(\text{core})}_{(d_x,d_z,h,w)} = \frac{w h (2d_z+d_x+1)(3d_x+1) + S_{\text{TRGP}} }{4 w h d_z d_x}.
\end{align}
In the cache, the routing cost adds an additional $1 + \frac{N_2-1}{N_2d_x} \sim 1 + \frac{1}{d_x}$ multiplier when using surface code patches of distance $d_x$ and $d_z$. Hence, the total routing costs including both the core and the cache is given by
\begin{align}
    O^{(\text{total})}_{(d_x,d_z,h,w)} = \frac{\tilde{O}^{(\text{core})}_{(d_x,d_z,h,w)} + d_z(N_2(d_x+1)-1))}{d_xd_z(4wh+N_2)},
    \label{eq:Ototal}
\end{align}
where we defined $\tilde{O}^{(\text{core})}_{(d_x,d_z,h,w)} = 4whd_xd_z O^{(\text{core})}_{(d_x,d_z,h,w)}$.

For the Hubbard model, the total number of logical qubits $N_{\text{TLQ}}$ which excludes those used in the magic state factory is 
\begin{align}
    N_{\text{TLQ}} = 4wh + N_2,
    \label{eq:NCC}
\end{align}
with
\begin{align}
    N_2 = 2L^2 + \frac{L^2}{2} + 3 - 4wh.
    \label{eq:N2comp}
\end{align}
since there are $4wh$ logical qubits in the core. The total number of physical qubits used in the algorithm is then 
\begin{align}
    N_{\text{phys}} = 2d_xd_z N_{\text{TLQ}}O^{(\text{total})}_{(d_x,d_z,h,w)},
    \label{eq:Nphys}
\end{align}
where we used the fact that a surface code patch can be realized using a rectangular region of size $2d_xd_z$.

We now compute the algorithm runtime. Let $\mu$ be the total number of injected magic states in the core and Pauli measurements in the algorithm. Recall that in \cref{subsec:RevPauliComp}, $\mu$ was shown to be given by $\mu = 4N_{\text{TOF}} + N_T$. The time $T_{b}$ required to inject magic states via lattice surgery is thus $T_{b} = \mu (d_m+1) T_{\text{surf}}$ where $T_{\text{surf}}$ is the time required to measure the stabilizers and reset ancillas during one surface code syndrome measurement cycle. Using \cref{eq:P110,eq:P110,eq:P0001}, the parameters $d_x$, $d_z$ and $d_m$ are chosen such that 
\begin{align}
\label{eq:MarTot}
0.01634 \mu d_x  \ell   (21.93 p)^{(d_m+1)/2}  & < \delta/3 \\
\label{eq:ZLTot}
0.03148 \mu N_{\text{TLQ}} d_m d_x (28.91 p)^{(d_z+1)/2} & < \delta/3 \\
\label{eq:XLTot}
0.0148 \mu  d_m \frac{FA}{d_x} (0.762 p)^{(d_x+1)/2} & < \delta / 3,
\end{align}
In \cref{eq:MarTot}, we pessimistically take $d_x \ell = (d_x+2+h(3d_x+1))(d_x+2+w(2d_z+d_x+1)) - 4whd_xd_z$, which is the full area of the routing space in the core. By doing so, we consider a worst case scenario where the full routing space is used to perform lattice surgery for each injected magic state. In \cref{eq:ZLTot}, we used $\mathbb{P}_{Z_L} = \mathbb{P}_{(1,1,0,0)}$ since the difference with $\mathbb{P}_{(0,1,1,0)}$ is negligible. We also ignore higher order contributions arising from lattice surgery for the reasons explained in \cref{sec:LatticeSurgeryDecoder}. Lastly, in \cref{eq:XLTot} we pessimistically set $FA = (d_x+2+h(3d_x+1))(d_x+2+w(2d_z+d_x+1))$ to be the full area in the core. Such an assignment is done to take into account the possibility that the full routing space can used when preforming lattice surgery after injecting a magic state. Such a scenario would lead to a large $d_z$ distance, which also includes contributions from the logical qubits.

In \cref{tab:HubbardModelOverhead} we provide overhead costs associated with performing a Hubbard model simulation of lattice size $L$. Given the chosen values of $h$ and $w$ for a unit cells in the core, we first compute the minimum required values of $d_x$, $d_z$ and $d_m$ by solving \cref{eq:MarTot,eq:ZLTot,eq:XLTot} with $\delta \sim 1 \%$. We then compute the required number of physical qubits using \cref{eq:Nphys} and give the number of logical qubits in the core and in the cache. The last column includes the multiplicative factor resulting from the routing overhead costs. As can be seen, having more logical qubits in the core relative to those in the cache can substantially increase routing overhead costs. 

Apart from the chosen value of $d_m$ which satisfies \cref{eq:MarTot}, the algorithm runtime will depend on several factors. The first factor is the ratio of logical qubits used in the cache and in the core. Such a ratio will affect how many times one needs to read from, and write to the cache during the algorithm runtime. The second factor involves whether multi-qubit Pauli operators with $Y$ terms are measured using our twist-free approach or with twists. Lastly, the third factor includes runtime savings which can be achieved using our temporal encoding scheme for fast lattice surgery. As such, a more careful analysis of the algorithm runtimes is left for future work. We also leave the inclusion of resource costs associated with the magic state factories to future work. However, from the results of Refs.\cite{litinski2019magic,chamberland2020building}, we expect contributions from the magic state factories to only have a mild effect on the total resource overhead costs shown in \cref{tab:HubbardModelOverhead}.

\newpage
\bibliography{LatticeSurgery}

\end{document}